\documentclass[%
 reprint,
 superscriptaddress,
nofootinbib,
nobibnotes,
 amsmath,amssymb,
 APS,
numerical
]{revtex4}

\usepackage{graphicx}% Include figure files
\usepackage{xcolor}
\usepackage{dcolumn}% Align table columns on decimal point
\usepackage{bm}% bold math
\usepackage{pbox}
\usepackage{enumitem}
\usepackage{caption}
\usepackage{graphicx}

\usepackage{epstopdf}
\usepackage{filecontents}
\newcommand{\eat}[1]{}
\usepackage{tabularx}
\usepackage{rotating}

   {\end{multicols}}
\makeatother

\begin{document}

\title{\textcolor{black}{\textbf{Thermal conductivity of the side ledge in aluminium electrolysis cells: compounds as a function of temperature and grain size. }}}%

\author{A\"{i}men E. Gheribi}
\affiliation{%
 CRCT - Polytechnique Montr\'{e}al, Box 6079, Station Downtown, Montr\'{e}al, Qc, Canada
}%
 \email{aimen.gheribi@polymtl.ca (corresponding author) 
}
\author{Patrice Chartrand}
\affiliation{%
 CRCT - Polytechnique Montr\'{e}al, Box 6079, Station Downtown, Montr\'{e}al, Qc, Canada
}%

\date{\today}

\begin{abstract}
In aluminium electrolysis cells, a ledge of frozen electrolyte is formed, attached to the sides of the cell. The control of the side ledge thickness is essential in ensuring a reasonable lifetime for the cells. Numerical modelling of the side ledge thickness requires an accurate knowledge of the thermal transport properties as a function of temperature. Unfortunately, there is a considerable lack of experimental data for the large majority of the phases constituting the side ledge. The aim of this work is to provide, for each phase possibly present in the side ledge, a formulation of the thermal conductivity as a function of both temperature and size. To achieve this, we consider reliable physical models linking the density of the lattice vibration energy and the phonon mean free path to key parameters: the high temperature limit of the Debye temperature and the G\"{u}neisen constant. These model parameters can be obtained by simultaneous fitting of (i) the heat capacity, (ii) the thermal expansion tensor coefficient and (iii) the adiabatic elastic constants, on relevant physical models. Where data is missing, first principles (\textit{ab initio}) calculations are utilised to determine directly the model parameters. For compounds for which data is available, the model's predictions are found to be in very good agreement with the reported experimental data.
 
\end{abstract}

\maketitle

%\tableofcontents

\section{Introduction}

The aluminium industry is a major contributor to the economy of Qu\'{e}bec as it represents 10$\%$ of its trade balance and employs over 10,000 people. Because of its desirable physical properties, aluminium is an attractive material for the future and a must for several economic sectors, such as transport, road infrastructure, packaging and construction. The production of Aluminium from its ores is carried out in a two-step process. First, the alumina (Al$_{2}$O$_{3}$) is produced from Bauxite following the Bayer process. Then, the alumina is reduced according to the so-called Hall--H\'{e}roult process. This process basically consists in dissolving alumina in a molten fluoride electrolyte, then the alumina is elctrolyzed between a consumable coke anode and liquid aluminium acting as a cathode, according to the global reaction:
\begin{equation}
2 Al_{2}O_{3} (_{dis.})+3C_{(anode)} \rightarrow 4 Al_{(liq.)} + 3CO_{2} \left( g \right)
\end{equation}
for which 12 electrons are exchanged. The chemical composition of a standard fluoride electrolyte melt is typically, in wt$\%$: 80 Na$_{3}$AlF$_{6}$ -- 12 AlF$_{3}$ -- 5 CaF$_{2}$ -- 3 Al$_{2}$O$_{3}$. This particular composition is commonly called cryolitic bath. Depending on the process optimization, the composition of the electrolytic melt could be subject to variations and could eventually contain lithium, potassium and magnesium additives. Usually, at the operating temperature, T$_{O}$, the electrolysis cell lies in the temperature range: $ T_{liq.}+5\,K \leq T_{O} \leq T_{liq}+15\, K $, where $T_{liq.}$ is the liquidus temperature. Consequently, a ledge of frozen electrolyte melt is formed, attaching itself to the sides of the cell.\\

 The control of the side ledge is essential in order to maintain a reasonable life span for the electrolysis cell. The ledge acts as a protective layer to prevent erosion and chemical attack on both cryolitic melt and liquid aluminium on the side wall material. Several studies, in particular those using finite element analysis, have been carried out to investigate the evolution of the side ledge thickness as a function of cryolitic melt composition and electrolyse cell parameters such as cell voltage, cell current, feed cycle, frequency and duration of anode effects, amount of alumina covering over the anodes, metal tapping, anode changes, etc.\\
 
 It is important to note that the characterization of the thermal transport properties of materials requires, strictly speaking, two independent properties among the following: thermal diffusivity $a$, thermal effusivity $e$, thermal conductivity $K$, volumetric heat capacity at constant pressure $C_{P}$ and  density $\rho$. The heat capacity at constant pressure can be measured accurately by traditional calorimetric methods such as drop calorimetry or by Differential Scanning Calorimeter (DSC). Although there exists experimental methods to directly measure the thermal conductivity (steady-state method) and effusivity (transient methods), the thermal diffusivity measured with laser flash method is now the standard method to characterize the thermal transport properties of materials. $K$ and $e$ are then deduced following the relationship: $K=a\times \rho \times C_{p}$ and $e=K/\sqrt{a}$. The lack of experimental data is, by far, more severe for thermal diffusivity than for heat capacity and density.\\  
 
 The only experimental data available in the literature related to thermal transport properties of the side ledge (S--L) in the Hall--H\'{e}roult cells is dated from 1970, by Haupin~\cite{Haup}. The thermal transport was measured using a hot wire technique apparatus~\cite{Haup2} and the results are presented as a function of temperature, from room temperature up to the liquidus temperature, as well as in function of alumina content. The following empirical expression of the thermal conductivity, as a function of temperature and alumina content, was then proposed by Haupin~\cite{Haup}: 
\begin{equation}
K_{S-L}\left(T,W_{Al_{2}O_{3}}\right)= 0.594+6.35\times 10^{-4}W_{Al_{2}O_{3}}+6.35\times 10^{-4}T \quad [W/(m.K)]
\end{equation}  
 where $W_{Al_{2}O_{3}}$ denotes the alumina weight fraction. However, such equation cannot be used as a standard value of the thermal conductivity of the side ledge. First the thermal conductivity of complex microstructures, such as the side ledge in aluminium electrolysis cells, depends strongly on their composition and properties. Here, the Haupin representation of the thermal conductivity only depends on the temperature and the alumina content, not on the amount of other phases which are generally present in the side ledge. Nowadays, in the aluminium electrolysis cells,  it is common to add, on top to AlF3, other additives like LiF to alter different properties (melting point, density, conductivity, etc.) of the electrolyte~\cite{Totti}. For an electrolyte containing LiF, the phases  present or potentially present in the side ledge are listed as~\cite{Char}: 

\begin{tabular}{p{6cm} p{6cm}}

\begin{enumerate}
  \item NaF (cubic) 
    \item LiF (cubic)
    \item  CaF$_{2}$ (cubic)
    \item  $\alpha$--Na$_{3}$AlF$_6$(monoclinic) 
    \item  $\beta$--Na$_{3}$AlF$_6$ (cubic)
    \item  Na$_{5}$Al$_{3}$F$_{14}$ (tetragonal)
    \item $\alpha$--NaCaAlF$_{6}$(monoclinic) 
    \item $\beta$--NaCaAlF$_{6}$ (trigonal)
    \item Na$_{2}$Ca$_{3}$Al$_{2}$F$_{14}$ (cubic)
    \item Na$_{4}$Ca$_{4}$Al$_{7}$F$_{33}$ (cubic)
\end{enumerate} & 
\begin{enumerate}
    \item $\alpha$--Na$_{2}$LiAlF$_{6}$(monoclinic)
    \item $\beta$--Na$_{2}$LiAlF$_{6}$ (cubic)
    \item LiCaAlF$_{6}$ (hexagonal)
    \item $\alpha$--Al$_{2}$O$_{3}$ (trigonal)
    \item $\gamma$--Al$_{2}$O$_{3}$(1)(cubic)
    \item $\gamma$--Al$_{2}$O$_{3}$(2)(tetragonal)
    \item NaAl$_{11}$O$_{17}$(hexagonal)
    \item Al$_{4}$C$_{3}$ (trigonal)
    \item  MgF$_{2}$ (tetragonal)
    \item Coke (anodic)
\end{enumerate}\\

\end{tabular}

The large number of phases which can potentially be present in the side ledge indicates that the thermal conductivity may vary by a factor up to 100--300$\%$, depending on the phases in equilibrium in the side ledge and their amounts. If the side ledge contains, for example, a large amount of NaF, its thermal conductivity should be much higher than if contained a large amount of NaCaAlF$_{6}$. In general, the phases composition depends on the thickness and height as they depend strongly on the temperature gradients. This makes the side ledge microstructures complex systems. In others words, the simplist Haupin formulation of the thermal conductivity (Eq.2) is not really appropriate to describe the thermal transport properties within the side ledge.\\

The thermal conductivity depends strongly on the microstucture and its proper parameters. By ``proper microstructure parameters'', we mean the classical: type of microstructure, average phases grain size, intrinsic grains porosity, inter grains porosity level, grain size distribution and grain orientation. Moreover, according to Haupin~\cite{Haup}, thermal conductivity increases with temperature, which is not consistent with a physical behaviour; the thermal conductivity of insulating materials are due to phonon--phonon scattering and should, on the opposite, decrease with 1/T~\cite{More}. This inconsistency is probably due to the experimental technique used. As discussed by DiGuilio and Teja~\cite{DeGui}, and later by Gheribi et al.~\cite{Gheribi8}, the Haupin formulation is only valid in a narrow temperature range near the liquidus temperature. \\

Without loss of generality, the thermal conductivity of a real ``micro-structure'', $K^{ms}$, consisting of $x$ multiple phases, can be expressed in the following form:
\begin{equation}\label{Eq3}
K^{ms} =\psi [\phi_{1}(\underline{x},T), \phi_{2}(\underline{x},T),\cdots,\phi_{n}(\underline{x},T),K_{1}(T,d_{1},p_{1}),K_{2}(T,d_{2},p_{2}), \cdots, K_{n}(T,d_{n},p_{n}),P]+\Psi^{ms}
\end{equation}
In this equation, $\underline{x}$ is the vector composition and T the average thermodynamic temperature. The two functions $\phi_{i}$ and $\lambda_{i}$ and the two variables $d_{i}$ and $p_{i}$ are respectively the volume fraction, the thermal conductivity, the average grain size and the average porosity of the i~$^{th}$ phase of the microstructure. $P$ is the total inter-grain porosity and $\Psi^{ms}$ is a function which takes into account all other microstructure parameters. The $\psi$ function is dependent only on the type of microstructure. An analytical expression for $\psi$ does not necessarily exists, it can be defined with functional equations or by ordinary differential equations. In addition to the relevant microstructure parameters, the information required for each phase in order to predict $K^{ms}$ is: (i) the volume fraction and (ii) the thermal conductivity. The thermal expansion coefficient of each phase gives the temperature dependence of the volume fraction. Unfortunately, neither the thermal conductivity nor the thermal expansion data is available for the vast majority of the phases constituting the side ledge. The list of these phases is given in section V.\\

To alleviate this lack data, we have developed, for electrically insulating material and semiconductors, a generalized self consistent method~\cite{Gheribi1,Gheribi2,Gheribi3,Gheribi4,Gheribi5} to predict the Gibbs free energy of solids from 0~K up to the melting temperature and the thermal conductivity from room temperature up to the melting point. The method combines the quasi-harmonic approximation for the density of the lattice vibration energy and from anharmonic Umklapp processes (U-process) for the phonon-phonon scattering at high temperature, approximately above one third of the Debye temperature. The temperature dependent thermal expansion and the bulk modulus are then derived from the Gibbs energy according the classical Maxwell relations. \\

The formalism contains a few key physical parameters, namely the Debye temperature ($\Theta_{D}$), the Gr\"{u}neisen constant ($\gamma$), the energy and entropy of formation of defect ($\Delta E_{\square}^{f}$ and $\Delta S_{\square}^{f}$ ). The originality of our methodology lies in the fact that the parameters are obtained by fitting simultaneously all the available experimental data on thermodynamic properties, which can be directly derived from Gibbs energy. The thermal conductivity is not taken into account when determining the parameters, it is fully predicted. In case of severe lack of data, i.e. if no reliable data is available for heat capacity, the model's parameters can be calculated, with an appreciable accuracy, using Density Functional Theory~(DFT)~(\textit{ab initio}) with suitable pseudopotential~\cite{Gheribi4,Gheribi5,Gheribi6,Gheribi7,Gheribi8}.   \\

 The aim of this work is primarily to provide, for each phase that can be present in the side ledge, an explicit formulation of the thermal conductivity as a function of both temperature and size, and also of the molar volume (density) as a function of temperature. This allows us to calculate, for a given side ledge microstructure, the phases volume fractions and therefore making possible the prediction of $K^{mis}$. \\
 
\section{ Modelling  }

\subsection{Lattice thermal conductivity model}
The differential form of Fourier's Law of thermal conduction states that for homogeneous material, there is a linear relationship between the local heat flux density, $\vec q$, and the temperature gradient~$\vec \nabla T $:

\begin{equation}
\vec q = - \underline{\underline{K}} \vec \nabla T
\end{equation}

Thermal conductivity is a second rank tensor and written as $ \underline{\underline{K}} $. Depending on the crystal symmetry, the number of independent components in $ \underline{\underline{K}} $ lies between a single one (for cubic symmetry) and up to 9 (for triclinic symmetry). For cubic, tetragonal, hexagonal, tetragonal and orthorhombic symmetries, $ \underline{\underline{K}} $ is diagonal, and for cubic systems, all the diagonal components are equal. For isotropic systems, the conventional thermal conductivity is given by the average quantity in the different directions:
\begin{equation}
K= \dfrac{1}{3}tr \left(\underline{\underline{K}} \right)
\end{equation}
From a practical point of view, the experimentally measured thermal conductivity is in fact the average thermal conductivity. This is primarily due to the experimental difficulties to measure the thermal conductivity tensor components and to prepare single crystal samples with an high level purity. As a result, very little data can be found in the literature.\\
For electrically insulating materials, thermal conduction comes from atomic vibrations, the so-called lattice thermal conductivity, and radiative heat transfer (if the medium is translucent): the so-called radiative thermal conductivity. Thermal conductivity is then written as a sum of these two contributions:
\begin{equation}
K = K^{lat.}+K^{rad.}
\end{equation}
Radiative thermal conductivity depends on the optical properties of the crystal: 
\begin{equation}
K^{rad.}\left(T\right)=\dfrac{16}{3}\dfrac{\sigma n^{2}}{\sigma_{e}}T^{3}
\end{equation}
where n is the refractive index,  $\sigma$ = 5.6704$ \times $ 10$^{-}$8 W/(m$^{2}$K$^{4}$) is the Stefan-Boltzman constant and $\sigma_{e}$ is the extinction coefficient. \\

The general approach to calculate the lattice thermal conductivity is to solve the Boltzmann transport  equation under the relaxation time approximation, while the scattering cross section is calculated by perturbation theory. In the continuum region, assuming (i) localized phonon modes, (ii) a Debye phonon density of state and (iii) that the contribution of optical phonons is neglected, the resolution of the Boltzaman thermal equation gives~\cite{Holl}:
\begin{equation}
K^{lat.}\left(\omega\right)=\frac{1}{2\pi}\int_{0}^{\omega_{D}}\dfrac{\tau\left(\omega\right)}{\nu_{g}}\omega^{2}C_{V}\left(\omega\right)d\omega
\end{equation}
where $\omega_{D}$ is the Debye frequency for an acoustic phonon branch, $C_{V}$ is the heat capacity per atom at constant volume, $\nu_{g}$ is the phonon group velocity and $\tau$ is the combined phonon single mode scattering relaxation time. At high temperature ($T \leq \Theta_{D}$), the lattice thermal conductivity decreases with temperature due to a large scattering of phonons through the Umklapp process, shortening the mean free path~\cite{Berm}. Considering the first order perturbation theory, Callaway~\cite{Cala,Cala2}, the phonon single mode scattering relaxation time of the U-processes $ \tau_{U} $ can be written as~\cite{Slack}:
\begin{equation}
\dfrac{1}{\tau_{U}\left(\omega\right)}=T\omega^{2}e^{-C/T}\Phi_{ph-ph}
\end{equation}
where $\Psi_{ph-ph}$ is a temperature and pressure dependent function describing the phonon-phonon scattering magnitude and $C$ is a constant. The integration of Eq.8 is then simple and gives:
\begin{equation}
K^{lat.}=\dfrac{C_{V}\omega_{D}^{ac.}}{2\pi^{2}\nu_{g}\Phi_{ph-ph}}
\end{equation}

where $ \omega_{D}^{ac.} = \omega_{D}n^{-1/3} $ is the acoustic Debye frequency linked to the Debye frequency through $n$, the number of ions per primitive cell. $n$ can be seen as a number describing the complexity of the crystal:  $n$ is larger as the crystal symmetry is more complex.  Several expressions of $\Phi_{ph-ph}$ were developed with different degrees of complexity. The most relevant formulation of $\Phi_{ph-ph}$ was originally proposed by Callaway~\cite{Cala,Cala2} and later by Slack~\cite{Slack2, Mingo} and relied only on a few key physical properties for its parametrization. Without going into details, which can be found in~\cite{Slack,Gheribi1, Gheribi3, Gheribi5}, Callaway and Salck's expression of $\Phi_{ph-ph}$, combined with Eq. 10, gives an expression which can be decomposed into two terms: a materials constants term ($Mat-Cst.$) and one describing the temperature dependence ($T-Dep.$):
\begin{equation}
K^{lat.}\left(T\right)=A \times\underbrace{\dfrac{\bar{m}\theta_{D}^3}{n^{2/3}}\times \dfrac{1}{\gamma_{\infty}^{2}-0.514\gamma_{\infty}+0.228}}_{Mat.-Cst.}\times\underbrace{\dfrac{\delta(T)e^{\phi/T}}{T}}_{T-Dep.}
\end{equation}  

The ($ \infty$) symbol in the Gr\"{u}neisen parameter indicates the ``\textit{high temperature limit}''. $A$ is a constant equal to $6.51982\times 10^{-3}\left(k_{b}/\bar{h}\right)^{3}$. $\bar{m}$ is the average mass per atom defined from the molecular mass of the compound, $\overline{M}$ as: $\bar{m}=\overline{M}/(n_{0}N_{A})$, ($N_{A}$ and $n_{0}$ are respectively the Avogadro constant and the number of atoms per chemical formula). In the last equation, $\delta^{3}$ represents the average volume per atom, and $\delta$ can be defined from the molar volume as follows: $\delta=\left(V_{m}/(n_{0}N_{A})\right)^{1/3}$. 
\begin{equation}
\delta\left(T\right)=\delta\left(T_{0}\right)\times exp\left[\int_{T_{0}}^{T} \dfrac{1}{3}\alpha_{V} \left(T\right)\right]
\end{equation}
$ \alpha_{V}=V_{m}^{-1}\left(\partial V_{m}/\partial T \right)_{P} $ is the volumetric thermal expansion. $T_{0}$ is a reference temperature which is assumed, for solids, to be equal to the standard temperature of 298.15 K.\\

 The value of the constant $\phi$ is not clearly defined. In fact, two values of $\phi$ can be found in the literature: either $\phi=0$ or $\phi=\Theta_{D}/3$. In other words the exponential term in Eq.~11 can appear or not. There is no criteria based on physical considerations to fix a value for $\phi$. From an empirical point of view, our research have shown that $\phi=\Theta_{D}/3$ is more suitable for Chlorides whereas $\phi=0$ is better suited for other ionic compounds. For all compounds composing the side ledge of aluminium electrolysis cells, fluorides, oxides and carbides, we assume that:
\begin{equation}
\phi^{Fluorides}=\phi^{Oxides}=\phi^{Carbides}=0
\end{equation} 
 \\
We now come to the description of the temperature-grain size dependence upon the thermal conductivity. Depending on the nature of the chemical interactions within the material, the thermal conductivity of polycrystalline materials could be considerably lower than that of its corresponding bulk crystals. This difference is due to the existence of grains, with known sizes and shapes, inducing thermal resistance at the grain boundaries. When heat is conducted from one grain to another, the temperature is not continuous at the boundary, there is a temperature jump, $\Delta T$, which is proportional to the heat flow ($\Delta \dot{Q} $): $\Delta T=R_{GB}\times \Delta\dot{Q}$ ($ R_{GB} $ being the thermal resistance across the grain boundary). We recently developed a formalism~\cite{Gheribi5}, showing a high predictive capacity, to take into account the grain size effect, and thus the presence of grain boundaries within the microstructure. In a nutshell, we derived an expression of thermal conductivity as a function of both temperature average grain size ($d$), as follows~\cite{Gheribi5}:
\begin{equation}
K^{lat.}\left(T,d\right)=K_{\infty}^{lat.}\left(T\right)\left[1- \sqrt{\dfrac{\sigma(T)}{d}}arctan\left(\sqrt{\dfrac{d}{\sigma(T)}}\right)\right]
\end{equation}
where $ K_{\infty}^{lat.}$ is the temperature dependent thermal conductivity of the single crystal and $\sigma$ is a characteristic length with the same physical meaning as the Kapitza length, $L_{K}$~\cite{Nan}. Briefly, the Kapitza length is defined as $ L_{K}=K_{\infty}^{lat.} R_{GB} $ which can then be seen as the average grain size, reducing the lattice thermal conductivity by half compared to a single crystal. For an average grain size larger than $100 L_{K}$, the grain boundaries have no contribution upon the thermal conductivity. We have shown that the characteristic length $\sigma$ is temperature dependent and can be expressed as a function of T:
\begin{equation}
\sigma\left(T\right)=\left[\dfrac{2\pi^{2}\nu_{g}\left(T\right)\hbar^{3}n}{C_{V}\left(T\right)k_{B}^{3}\theta_{D}^3}\right]
\end{equation}
where $\nu_{g}$ and $C_{V}$ are not the usual thermophysical properties which are experimentally measured and available in the literature. Indeed, in the literature, it is more common to find the adiabatic bulk modulus and the $C_{P}$.  For isotropic materials, the phonon group velocity is linked to the adiabatic bulk modulus, through the well known relation:
\begin{equation}
\nu_{g}\left(T\right)=\sqrt{\dfrac{B_{S}\left(T\right)V_{m}\left(T\right)}{\overline{M}}}
\end{equation}
and $C_{V}$ is linked to $C_{P}$ via:
\begin{equation}
C_{V}\left(T\right)=\dfrac{C_{P}\left(T\right)}{\left[1+\alpha\left(T\right) \gamma_{\infty}   T\right]}
\end{equation}\\
In summary, it is therefore possible to predict the thermal conductivity of polycrystalline insulating materials as a function of both temperature and grain size, from an accurate knowledge of two physical parameters, $\Theta_{D}$ and $\gamma_{\infty}$, and three temperature dependent thermodynamic properties: $C_{P}$, $\alpha_{V}$ and $B_{S}$ which can also be described by $\Theta_{D}$ and $\gamma_{\infty}$. Thus, few key physical properties can describe simultaneously the lattice thermal conductivity as a function of both temperature and grain size.

\subsection{Thermodynamics modelling}

Usually, in the literature, the thermodynamic properties of compounds are formulated in the form of empirical functions containing a large number of parameters given in different temperature ranges~\cite{CALP}. However, such representations are not based on physical models and require a huge amount of experimental data in a wide range of temperatures, generally in the domain where the considered phase is stable. From a theoretical point of view, the thermodynamic properties of materials are a sum of several physical contributions, for example lattice vibration, electronic, magnetic and defects. Each physical contributions may be described by few key physical parameters if suitable physical models are considered. In the present situation, both thermodynamic and thermal transport properties result from the lattice vibration and possibly the defect contribution at high temperature. In a recent works~\cite{Gheribi4,Gheribi6,Gheribi7}, a thermodynamically self consistent (TSC) method has been developed to predict thermodynamic properties of a large variety of materials as a function of temperature and pressure. Here it is not necessary to detail the TSC method as this is already published~\cite{Gheribi4,Gheribi6,Gheribi7}. Briefly, the TSC method extends the quasi-harmonic approximation (QHA) model~\cite{Blan} by incorporating a minimization procedure to ensure the conservation of Maxwell relations. Considering an isotropic or cubic solid, the TSC method allows to calculate its molar volume, isothermal and adiabatic bulk modulus, Debye temperature Gr\"{u}neisen parameter, volumetric thermal expansion, heat capacities at constant pressure and volume. The lattice vibration contribution to heat capacity at constant volume is expressed as:
\begin{equation}
\begin{aligned}
C_{V}\left(T\right)=9n_{0}R\left(\dfrac{T}{\theta_{D}}\right)^{3}\left[4\int_{0}^{x_{D}}\dfrac{x^{3}}{e^{x}-1}dx -\dfrac{\left( \theta_{D}/4\right)^{4}}{e^{\left( \theta_{D}/4\right)}-1}\right]
\end{aligned}
\end{equation}

$n_{0}$ is the number of atoms per formula units and $x_{D}=\frac{\theta_{D}}{T}$ defines the temperature dependence of the 
Debye frequency. Thereafter, the temperature dependence of the isothermal bulk modulus is given by: 
\begin{equation}
B_{T}\left(T\right)=B_{0}+\dfrac{\gamma_{\infty}}{V_{m,0}}\left(\dfrac{\partial B}{\partial P}\right)_{T}\int_{0}^{T} C_{V}\left(T'\right)dT'
\end{equation} 

a $0$ index means that the property is considered at 0 K, $B_{0}$ is the adiabatic or isothermal (they are identical) bulk modulus at 0 K . Then the volumetric thermal expansion is deduced from the definition of the Gr\"{u}eisen parameter by solving the following differential equation:
\begin{equation}
\alpha_{V}\left( T\right)e^{\int_{T_{0}}^{T}\alpha_{V}\left( T'\right)dT'} =\gamma_{\infty} \dfrac{ C_{V}\left(T\right)}{ V_{m}\left(T_{0}\right) B_{T}\left(T\right)}
\end{equation}

 Note that the left side of Eq.20 defines the temperature dependence of the molar volume above $T_{0}$. Thereafter, the heat capacity at constant pressure and the adiabatic bulk modulus are respectively expressed, as a function of temperature, as:

\begin{equation}
 C_{P}\left(T\right)= C_{V}\left(T\right) \left[ 1+\alpha\left(T\right) \gamma_{\infty} T \right]
\end{equation}  
and 
\begin{equation}
 B_{S}\left(T\right)= B_{T}\left(T\right) \left[ 1+\alpha\left(T\right) \gamma_{\infty} T \right]
\end{equation}  

Among the other physical contributions which can influence the total energy, in addition to vibrational contributions, only the thermal defect contribution could be significant for the phases potentially present in the side ledge. The thermal defect contribution to the heat capacity and thermal expansion obeys in general to an Arrhenius-like behaviour~\cite{Grim}: 
\begin{equation}
C_{P}^{def.}\left(T\right)=n_{0}R\left( \dfrac{\Delta E_{def.}}{RT}\right)^{2} e^{-\Delta G_{def.}/RT}
\end{equation}
and
\begin{equation}
\alpha_{V}^{def.}\left(T\right)=\left( \dfrac{\Delta V_{ def.}}{\delta}\right)\left( \dfrac{\Delta E_{def.}}{RT^{2}}\right) e^{-\Delta G_{def.}/RT}
\end{equation}

where $\Delta V_{def.}$ and $\Delta E_{def.}$ are respectively the volume and the energy of formation of thermal defect and the Gibbs energy of formation of thermal defect is defined as: $\Delta G_{def} =\Delta E_{def}-T \Delta S_{def.}$. In general, it is difficult to experimentally determine both $\Delta V_{def.}$ and $\Delta S_{def}$, it is assumed that:  $\left(\frac{\Delta V_{ def.}}{\delta}\right)\simeq \frac{1}{2} $ and $S_{def.}=3n_{0}R$~\cite{Ceza} and the energy of formation of defect is determined by fitting the heat capacity at high temperature, above about 0.7--0.75 of the melting temperature. When direct measurements of $\Delta E_{def}$, e.g. by positron annihilation method, are available together with  reliable experimental data at high temperature for both heat capacity and thermal expansion, it is then possible to adjust the generic values of $\Delta S_{def.}$ and $\Delta V_{def.}$.

\section{Thermal conductivity and thermodynamic model parametrization}

In addition to the 3 parameters describing the cold energy cure $E\left(V\right)$ , i.e. $V_{m,0}$, $B_{0}$ and $\left(\partial B/\partial P\right)_{T}$, the two key physical parameters which describe the thermal transport and the thermodynamics properties are: $\Theta_{D}$ and $\gamma$. If sufficient experimental data is available for $B_{S}$, $\alpha_{V}$ and $C_{P}$ in a wide range of temperature, ideally from 0 K up to the melting temperature, the model parameters can be determined by a simultaneous fitting of these property on the functions described above, according to the methodology established by Gheribi and Chartrand in a prior work~\cite{Gheribi1}. For many phases, in fact for the large majority of them, there is a considerable lack of data. One exception: simple compounds (NaF, CaF$_{2}$,LiF, $\alpha$--Al$_{2}$O$_{3}$) for which sufficient experimental data is available for $B_{S}$, $\alpha_{V}$ and $C_{P}$. For every other compound present in the side ledge, experimental data is missing or only partially available. To overcome such a severe lack of data, DFT calculations combined with the TSC method can play an important role in making accurate predictions for both thermodynamic and thermal transport properties, as previously demonstrated by Gheribi et al.~\cite{Gheribi4}. \\

 The Debye temperature of an isotropic material is defined from ground state properties according to the following relationship~\cite{Poire,Blan,Xiao}: 
\begin{equation}\label{eq:MTheta}
\theta_{D}\left(V_{m}\right)=\left(\dfrac{\hbar}{k_{B}} \right)\left( 6 \pi^{2} n_{0} N_{A}\right)^{1/3}V_{m}^{1/6}f\left(\nu\right)\sqrt{\dfrac{B_{0}}{\overline{M}}}
\end{equation}
 $f\left(\nu\right)$ is a function depending on the Poisson ratio and given by:
 
 \begin{equation}
  f\left(\nu\right)=\left\lbrace  3 \left[ 2 \left( \dfrac{2}{3} \dfrac{1+\nu}{1-2\nu}\right)^{3/2}+ \left( \dfrac{1}{3} \dfrac{1+\nu}{1-\nu}\right)^{3/2} \right]^{-1} \right\rbrace  ^{1/3}
 \end{equation}

The Poisson ratio of oxides and fluorides is typically in the range of 0.2 to 0.3 and is, in general, 0.25~\cite{TMLP}. As proposed by Toher et al.~\cite{PRB}, a constant value of 0.25 can be assumed for $\nu$, at least for oxides and fluoride. Consequently, we assume that $ f\left(\nu\right)\simeq f\left(0.25\right)\simeq 0.86 $ for the compounds.
The procedure to calculate $B_{0}$ by DFT is well established. Basically, it consists in calculating the energy difference between the equilibrium lattice~\textbf{R} and an expanded or compressed lattice (\textbf{R$'$}), by applying very small strains ($\epsilon$) in each crystallographic direction (in order to make sure we stay within the elastic domain). $B_{0}$ is then obtained by fitting the energy curve $E\left(\epsilon\right)$ by a suitable equation of state (EOS). In this work, we have considered the Birch-Murnaghan EOS, derived from finite strain theory, as the applied deformation is relatively weak~\cite{Poire}. This procedure is then repeated under external pressures of 1, 2, 3 and 5 GPa in order to obtain the Gr\"{u}neisen parameter. Indeed, assuming that the volume dependence of all modes of the phonon frequencies is identical to the volume dependence of the Debye frequency, the Gr\"{u}neisen parameter is defined as:
\begin{equation}\label{eq:MTheta}
\gamma_{\infty}=-\dfrac{\partial \ln\left[\theta_{D}(V_{m}) \right] }{\partial \ln V_{m}}
\end{equation}  \\
The DFT calculations are based on the Plane-Wave basis sets and are done using the Vienna ab initio Simulation Package (VASP)~\cite{2,3,4,5}. The Projected Augmented Wave (PAW) approach is employed to represent the core electrons~\cite{6,7}. Generalized Gradient Approximation (GGA) parameterized by Perdew, Burke and Ernzerhof (PBE)~\cite{8,9} was used as the exchange-correlation functional. Plane-Wave kinetic cut-off energy of 520 eV and Monkhorst Pack grid of suitable dimension (given in Table 1) to sample the Brillouin zone with a first order Methfessel-Paxton smearing parameter $\sigma$ of 0.02 eV are used to ensure the force and energy convergence criterion are better than 0.02 eV/${\buildrel _{\circ} \over {\mathrm{A}}}$ and 0.01 meV, respectively. \\

 In summary, the strategy employed to determine the model parameters is as follows:
 
 \begin{itemize}
 \item Method ``Exp.'' - For compounds for which sufficient experimental data is available in the literature in a wide range of temperatures for $B_{S}$, $\alpha_{V}$ and $C_{P}$: the parameters are determined by a simultaneous fitting of these properties on Eq .17 to Eq. 24. A temperature dependent thermal conductivity is then deduced via Eq. 11. 
 
 \item Method ``Mixed'' - For compounds for which few experimental data is available, only available in a narrow range of temperature or just for one property among $B_{S}$, $\alpha_{V}$ and $C_{P}$,  the experimental data is not completely avoided, in fact the DFT data is used as ``target parameters values'' for the theoretical model and are then adjusted to reproduce the experimental data. The temperature dependent thermal conductivity is then deduced via Eq. 11 as well as the thermodynamic properties for which data is not available by the corresponding equation given above.
 
 \item Method ``DFT'' - When no experimental data is available, except the structure and the lattice constants, the parameters are purely predicted by DFT calculations and the temperature dependent thermophysical and thermal transport properties are purely predicted.
 \end{itemize}

Note that the set $\left( B_{S}, \alpha_{V}, C_{P} \right)$ has been chosen to determine the model parameters because it corresponds to the properties which are measured directly by experiments. This allows a better accuracy in the determination of model parameters as only the experimental error of each property influences the optimization procedure.

\newpage 

\begin{sidewaystable}
\centering
\begin{tabular*}{\textwidth}{ c  @{\extracolsep{\fill}} c c c c c c c c c c c c c c c c c}
\hline
& NaF& CaF$_{2}$&LiF&$\alpha$--Na$_{3}$AlF$_6$ & $\beta$--Na$_{3}$AlF$_6$ & Na$_{5}$Al$_{3}$F$_{14}$ & $\alpha$--NaCaAlF$_{6}$ & $\beta$--NaCaAlF$_{6}$& Na$_{2}$Ca$_{3}$Al$_{2}$F$_{14}$&Na$_{4}$Ca$_{4}$Al$_{7}$F$_{33}$\\ \hline
Method& Exp. & Exp. &Exp. & Mixed&Mixed&Mixed*&DFT&DFT&DFT&DFT	 \\
Crystal System& Cubic & Cubic &Cubic & Monoclinic&Cubic&Tetragonal&Monoclinic &Trigonal& Cubic&Cubic	 \\
Space group&	Fm$\bar{3}$m&	Fm$\bar{3}$m&	Fm$\bar{3}$m&	P2$_{1}$/n&Fm$\bar{3}$m&P4/mnc&P2$_{1}$/c&P321&I2$_{1}$3&Im$\bar{3}m$\\
$n$&	2&	3&	2&	20&40&44&72&27&42&96	\\
k-points grid &	----&	---&---&	9$\times$9$\times$9&9$\times$9$\times$9&9$\times$9$\times$3&10$\times$15$\times$5&10$\times$10$\times$15&5$\times$5$\times$5&5$\times$5$\times$5\\
$\rho_{298}$(kg/m$^{3}$)&	2802&	3180&	2639& 2885 & 2853&2995&2917&2877& 3002 & 2830 \\
$\Theta_{D}$(K)&	497&	495&	630&	515&	515&495&527&500&505&521\\ 
$\gamma_{\infty}$&	1.57&	2.10&	1.70&	1.61&	1.51&1.58&1.49&1.48&1.41&1.57\\ 
$\Delta E_{def.}$(kJ/mol)&	122.0&	168.3&	110.4&	---&---&---&---&---&---&---	\\ 
$\alpha_{0}$ (/K)&	6.27$\times$10$^{-6}$&	4.86$\times$10$^{-5}$&	-7.40$\times$10$^{-4}$&	1.28$\times$10$^{-4}$&1.30$\times$10$^{-4}$&7.54$\times$10$^{-5}$&6.65$\times$10$^{-5}$&6.15$\times$10$^{-5}$&6.27$\times$10$^{-5}$&7.19$\times$10$^{-5}$	\\ 
$\alpha_{1}$(/K$^{2}$)&1.43$\times$10$^{-7}$&	2.92$\times$10$^{-8}$&	2.10$\times$10$^{-7}$&	3.40$\times$10$^{-8}$&3.38$\times$10$^{-8}$&1.32$\times$10$^{-8}$&8.91$\times$10$^{-9}$&1.00$\times$10$^{-8}$&8.91$\times$10$^{-9}$&1.09$\times$10$^{-8}$	\\ 
$\alpha_{2}$&	1.66$\times$10$^{-2}$&	2.41$\times$10$^{-3}$&	9.98$\times$10$^{-2}$&	4.93$\times$10$^{-3}$&5.02$\times$10$^{-3}$&-7.53$\times$10$^{-4}$&-5.70$\times$10$^{-4}$&-5.70$\times$10$^{-4}$&-6.77$\times$10$^{-4}$	&-7.18$\times$10$^{-4}$		\\ 
$\alpha_{3}$(K)&	---&	---&	1.64$\times$10$^{1}$&1.64&1.68&-6.76$\times$10$^{-1}$&-6.27$\times$10$^{-1}$&-6.27$\times$10$^{-1}$&-5.76$\times$10$^{-1}$&-7.17$\times$10$^{-1}$	\\ 
$R_{\infty,0}^{lat.}$[(m$^{2}$K)/W]&	2.44$\times$10$^{-4}$&3.25$\times$10$^{-3}$&	-2.04$\times$10$^{-3}$&	3.51$\times$10$^{-3}$&4.84$\times$10$^{-3}$&3.50$\times$10$^{-2}$&9.28$\times$10$^{-3}$&2.36$\times$10$^{-2}$&-2.09$\times$10$^{-3}$&-3.66$\times$10$^{-3}$	\\
$R_{\infty,1}^{lat.}$[m$^{2}$/W]&	2.00$\times$10$^{-4}$&	3.95$\times$10$^{-4}$&2.43$\times$10$^{-4}$	&9.55$\times$10$^{-4}$	&1.32$\times$10$^{-3}$&1.65$\times$10$^{-3}$&1.59$\times$10$^{-3}$&1.17$\times$10$^{-3}$&1.08$\times$10$^{-3}$&2.28$\times$10$^{-3}$	\\
$R_{\infty,2}^{lat.}$[m$^{2}$/(W.K)]&	---&	---&	-3.72$\times$10$^{-8}$&	-5.39$\times$10$^{-8}$&-7.44$\times$10$^{-8}$&---&-3.28$\times$10$^{-7}$&-2.30$\times$10$^{-7}$&-4.84$\times$10$^{-8}$&-6.47$\times$10$^{-8}$	\\ \hline

&$\alpha$--Na$_{2}$LiAlF$_{6}$&$\beta$--Na$_{2}$LiAlF$_{6}$&LiCaAlF$_{6}$&$\alpha$--Al$_{2}$O$_{3}$&$\gamma$--Al$_{2}$O$_{3}$(1)&$\gamma$--Al$_{2}$O$_{3}$(2)& NaAl$_{11}$O$_{17}$&Al$_{4}$C$_{3}$&MgF$_{2}$&$\beta$-Si$_{3}$N$_{4}$	\\ \hline
Method&DFT&DFT&Mixed&Exp.&DFT&DFT&DFT&Mixed&Exp.&DFT.\\
Crystal System& Monoclinic&Cubic&Hexagonal&Trigonal&Cubic(Spinel)&Tetragonal&Hexagonal&Trigonal &Tetragonal&Hexagonal\\
Space group&	P2$_{1}$/n&Fm$\bar{3}$m&P$\bar{3}$1c&R$\bar{3}$C&Fd$\bar{3}$m&I4$_{1}$/amd&P6$_{3}$/mmc&R$\bar{3}$m&P4$_{2}$/mnm&P6$_{3}$/m		\\
$n$&20&40&18&10&10&10&58&7&6&14		\\
k-points grid &7$\times$7$\times$7&7$\times$7$\times$7&12$\times$12$\times$8&---&11$\times$11$\times$11&8$\times$8$\times$4&8$\times$8$\times$4&16$\times$16$\times$4&---&8$\times$8$\times$12&	\\
$\rho_{298}$(kg/m$^{3}$)&3030&3015&2968&3987&3651&3670&3099&2969&3176&3187	\\
$\Theta_{D}$(K)&452&452&525&920&880&880&926&960&595&975	\\ 
$\gamma_{\infty}$&	1.72&1.82&1.48&1.54&0.95&0.99&1.47&0.78&1.57&0.74\\ 
$\Delta E_{def.}$(kJ/mol)&---&---&---&---&	---&---&	---&---& ---&---&		\\  
$\alpha_{0}$ (/K)& 6.90$\times$10$^{-5}$&6.90$\times$10$^{-5}$& 5.14$\times$10$^{-5}$&4.19$\times$10$^{-5}$&2.02$\times$10$^{-5}$&2.02$\times$10$^{-5}$&3.36$\times$10$^{-5}$&2.58$\times$10$^{-5}$&1.92$\times$10$^{-5}$&1.66$\times$10$^{-5}$	\\ 
$\alpha_{1}$(/K$^{2}$)& 1.07$\times$10$^{-8}$&1.07$\times$10$^{-8}$& 1.04$\times$10$^{-8}$& -8.00$\times$10$^{-10}$&2.22$\times$10$^{-9}$&2.22$\times$10$^{-9}$&9.45$\times$10$^{-10}$&5.06$\times$10$^{-10}$&4.17$\times$10$^{-8}$&-1.82$\times$10$^{-9}$	\\ 
$\alpha_{2}$& -3.00$\times$10$^{-4}$&-3.00$\times$10$^{-4}$& -5.70$\times$10$^{-4}$&-1.14$\times$10$^{-2}$&-1.12$\times$10$^{-3}$&-1.12$\times$10$^{-3}$&-2.84$\times$10$^{-3}$&-2.02$\times$10$^{-3}$&---&-4.26$\times$10$^{-3}$			\\ 
$\alpha_{3}$(K)& -1.78&-1.78& -6.27$\times$10$^{-1}$&1.06&-3.12$\times$10$^{-1}$&-3.12$\times$10$^{-1}$&-3.65$\times$10$^{-1}$&-3.41$\times$10$^{-1}$&---&1.68	\\ 
$R_{\infty,0}^{lat.}$[(m$^{2}$K)/W]& 2.20$\times$10$^{-3}$&1.39$\times$10$^{-3}$&1.44$\times$10$^{-2}$& 8.85$\times$10$^{-4}$&2.99$\times$10$^{-4}$&2.67$\times$10$^{-4}$&-2.33$\times$10$^{-3}$&8.11$\times$10$^{-5}$&4.32$\times$10$^{-3}$	\\
$R_{\infty,1}^{lat.}$[m$^{2}$/W]& --& --&6.70$\times$10$^{-4}$&	1.05$\times$10$^{-4}$&4.15$\times$10$^{-5}$&4.51$\times$10$^{-5}$&3.19$\times$10$^{-4}$&1.57$\times$10$^{-5}$&2.76$\times$10$^{-4}$\\
$R_{\infty,2}^{lat.}$[m$^{2}$/(W.K)]&---& ---&---&---&--&---&5.99$\times$10$^{-8}$&---&---\\ \hline

\end{tabular*} 
\caption{Model parameters for the thermal conductivity and density as a function of temperature for phases involved or potentially involved in the side ledge of aluminium electrolysis cells. For each compound, the method used for the determination of the parameters, the crystal symmetry, the space group and the number of atom per primitive cell ($n$) are also specified and they are from: \cite{TMLP}. In addition to the model parameters, the empirical representations of both the lattice thermal conductivity and the volumetric thermal expansion as a function of temperature are also reported through the coefficients of the  $R_{\infty,i}^{lat.}$'s and $\alpha_{i}$(K)'s in accordance to Eq. 29 and Eq. 30.}
\end{sidewaystable}

\newpage
\section{Results and discussion}

 From a practical point of view, given the severe lack of data, the thermal conductivity of the side ledge formed in the aluminium electrolysis cells may be accurately estimated via Eq.3, given that the following theoretical and experimental informations are available: 
\begin{enumerate}
\item  \textbf{\underline{The chemical composition}}  of phases (compounds and solutions) of the side ledge.
\item \textbf{\underline{An accurate formulation}} of the thermal conductivity of \textbf{\underline{all phases}} present in the side ledge as a function of temperature.
\item \textbf{\underline{The volume fraction}} of each phase present in the side ledge.
\item \textbf{\underline{Accurate relevant informations}} on the microstructure: primarily the grain size, the porosity and the type of microstructure. Other informations such as the grain size distribution, the grain orientation and the chemical composition of the grain boundaries can also increase the accuracy of the model's predictions.     
\end{enumerate}

The present paper focuses on individual phases present or potentially present in the side ledge, i.e items 2 and 3. The chemical composition of the side ledge and the microstructural aspect of the thermal transport properties will be discussed in a future paper. In this work, two types of information are provided: the density and the thermal conductivity of each phase as a function of temperature. The density as a function of temperature being defined as:
\begin{equation}
\rho\left(T\right)=\rho_{298} e^{\int_{298}^{T}\alpha_{V}\left(T\right) dT}
\end{equation}

 where $ \rho_{298} $ is the density at the standard temperature of T = 298 K. \\
 
 The phases present in the side-ledge are listed in Table I, together with the corresponding parameters for the theoretical models for lattice thermal conductivity and volumetric thermal expansion. For practical reasons, above 298 K, the thermal expansions and thermal conductivities are in a clearer form: 
\begin{equation}
\alpha_{V}\left(T\right)=\alpha_{0}+\alpha_{1}T+\dfrac{\alpha_{2}}{T}++\dfrac{\alpha_{2}}{T^{2}}\qquad for \quad T\geq 298 \: K
\end{equation}
and 
\begin{equation}
K_{\infty}^{lat.}\left(T\right)=\dfrac{1}{R_{\infty,0}^{lat.}+R_{\infty,1}^{lat.}T+R_{\infty,2}^{lat.}T^{2}}\quad for \quad T\geq 298 \: K
\end{equation}

The coefficients of Eq. 29 and Eq. 30 are obtained by a least-square fitting procedure on the theoretical values above 298 K and up to the melting temperature. For each compound, the method used to determine the model parameters is indicated as: (i) ``Exp.'' when they were determined by a simultaneous fitting of experimental data of $\alpha_{V}$, $C_{P}$ and $B_{S}$ versus temperature, (ii) ``Mixed'' when the parameters are first determined by DFT and then adjusted to reproduce the experimental data on the available properties and (iii) ``DFT'' when the parameters and thus the thermal conductivity is purely predictive as no experimental data is available for $\alpha_{V}$, $C_{P}$ and $B_{S}$. The structure, i.e. the number of atoms per primitive cell ($n$) are from The Material Project~\cite{TMLP}. In addition to the phases constituting the side ledge, the properties of $\beta$--Si$_{3}$N$_{4}$ as it is, with silicon carbide (SiC), the main constituent of the side wall in the aluminium electrolysis cells. The side wall is directly facing the the side ledge, an accurate knowledge of its thermal transport properties is also highly desirable for  the heat fluxes balance across the side ledge. The thermal conductivity and thermodynamic properties of SiC were already formulated, using the ``Exp.'' method, in our previous work \cite{Gheribi1}.\\

For many phases present in the side ledge, there is no experimental data available allowing the characterisation of the thermal transport properties (thermal diffusivity, thermal conductivity or thermal effusivity). It is thus essential to demonstrate: (i) the reliability of the theoretical model (Eq. 11) for fluorides ans oxides as well as (ii) the DFT capability for an accurate prediction of the two key parameters $\Theta_{D}$ and  $\gamma$. Although the high predictive capacity of the theoretical model when predicting the lattice thermal conductivity, and the good accuracy of DFT in predicting the lattice vibrational properties for different type of compounds were already demonstrated in our previous works~\cite{Gheribi1, Gheribi2, Gheribi3, Gheribi4, Gheribi6, Gheribi7}, in this section, we aim to demonstrate as much as possible the accuracy of the proposed approach for the compounds constituting the side ledge.\\
\begin{figure}
\centering
\includegraphics[scale=0.32]{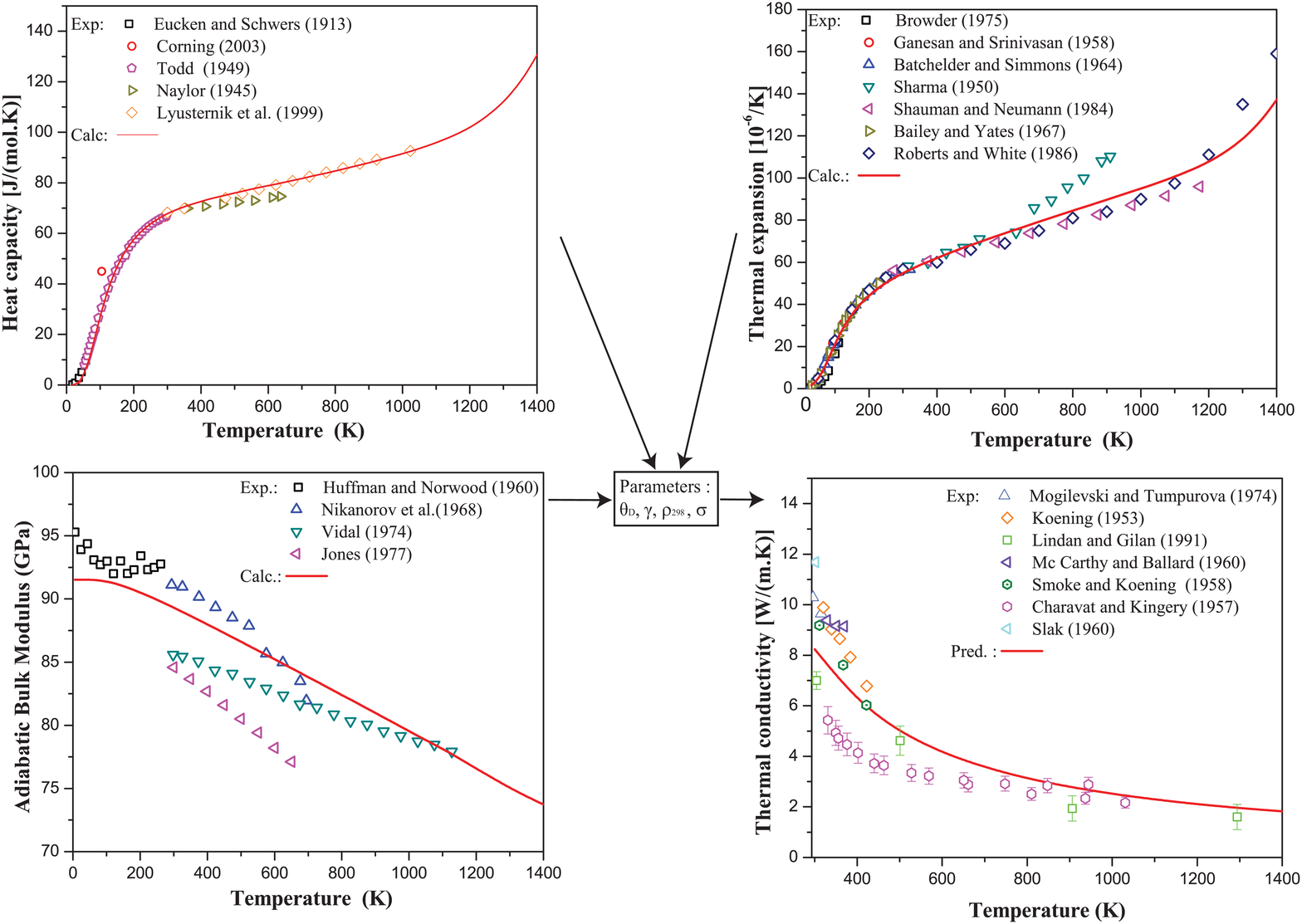} 
\caption{Illustration of method ``Exp.'' for the prediction of the CaF$_{2}$ lattice thermal conductivity (cubic,Fm$\bar{3}$m) as a function of temperature. The model parameters for the temperature and size dependent lattice thermal conductivity, $\rho_{298} $, $\Theta_{D}$, $\gamma$ and $\sigma$ are obtained by fitting simultaneously the heat capacity, the thermal expansion and the adiabatic bulk modulus as a function of temperature. The experimental data is referenced as follows: Eucken and Schwers (1913):~\cite{Euck}, Corning (2013):~\cite{Corn}, Todd (1949):~\cite{Todd}, Naylor (1945):~\cite{Nayl}, Lyusternik et al. (1999):~\cite{Lyus}, Browder (1975):~\cite{Brow}, Ganesan and Srinivasan (1958):~\cite{Gane}, Batchelder and Simmons (1964):~\cite{Batc}, Sharma (1950):~\cite{Shar}, Shauman and Neumann (1984):~\cite{Shau}, Bailey and Yates (1967):~\cite{Bail}, Roberts and White (1986):~\cite{Robe}, Huffman and Norwood (1960):~\cite{Huff}, Nikamorov et al. (1968):\cite{Nika}, Vidal (1974):~\cite{Vida}, Jones (1977):~\cite{Jone}, Mogilevski and Tumpurova (1974):~\cite{Mogi}, Koening (1953):~\cite{Koen}, Lindan and Gilan (1991):~\cite{Lind}, McCarty and Ballard (1960):~\cite{Maca}, Somoke and Koening (1958):~\cite{Smok}, Charavat and Kingery (1957):~\cite{Chara}, Slak (1960):~\cite{Slac60} }
\label{fig1a}
\end{figure}
The method ``Exp.'' for the prediction of the lattice thermal conductivity is illustrated in Figure 1 for CaF$_{2}$ (cubic,Fm$\bar{3}$m). This method is detailed extensively in our previous papers~\cite{Gheribi1, Gheribi2,Gheribi3,Gheribi4}, here we provide only the main principles. \\

It has already been demonstrated that for NaF, LiF, $\alpha$--Al$_{3}$O$_{3}$, MgO, $\alpha$--SiC~\cite{Gheribi1} NaCl, KCl~\cite{Gheribi3, Gheribi4} Li$_{2}$0~\cite{Gheribi5} the method ``Exp.'' shows a very good predictive capacity. The case of CaF$_{2}$ is interesting because of the dispersion of the experimental data for both adiabatic bulk modulus and thermal conductivity and the lack of experimental data for the heat capacity at high temperature. The advantages of the present simultaneous optimization procedure is that the temperature dependence of either $ C_{P} $, $\alpha_{V}$ and $B_{S}$ are correlated and thus even when the data is either not available, or just available in a narrow range of temperature, or scattered, a temperature dependence of the these quantities can be predicted or assessed from each other. The temperature dependence of the adiabatic bulk modulus of CaF$_{2}$ was assessed without examining the available experimental data or rejecting arbitrarily a given set of data. The order/disorder transition (from fluorite to anti-fluorite structure) is observed in thermal expansion of CaF2 very close to the allotropic transition (1423~K~\cite{Caf2}). The corresponding experimental heat capacity values do not show any significant effect of the vacancy formation, due to a lack of data above 1100~K. In the present work, vacancy formation contribution was deduced from the thermal expansion data, and the same contribution was used for the heat capacity. The experimental thermal conductivity data is also scattered at low temperature. So much so that a maximum of 100$\%$ discrepancy is observed between the largest~\cite{Slac60} and the lowest~\cite{Chara} reported data point. At first glance, it is difficult to discuss the reliability of the theoretical model's predictions. However, the theoretical predictions are in good agreement (less than 10$\%$ difference) with the most recent experimental data set (1991) reported by Lindan and Gilan~\cite{Lind}. Indeed, It is well known that for thermal diffusivity and thermal conductivity, recent experimental techniques provide, in general, more reliability than older ones. In absolute terms, our predictions are satisfactory, as the agreement with the overall sets of experimental data is good.     \\

We now come to discuss the most important aspect of the present section. It is crucial to demonstrate that both methods, ``Mixed'' and  ``DFT'', could provide accurate predictions for fluorides and oxides, as the thermal conductivity of the large majority of compounds present in the side ledge are predicted. In a recent study~\cite{arxiv}, for NaF, $\beta$--Na$_{3}$AlF$_{6}$ and Na$_{5}$Al$_{3}$F$_{14}$ we have performed a series of Equilibrium Molecular Dynamics (EMD) simulations in order to determine the thermal conductivity as a function of temperature. The predicted values of thermal conductivity obtained by EMD were then compared with those obtained by the ``Mixed'' method. The agreement between the theory and EMD simulations was found to be very satisfactory which confirmed the reliability of both the theoretical model given by Eq. 11 and the ``Mixed'' method for predicting the lattice thermal conductivity in the case of complex fluoride compounds. The aim of the present study is rather to validate the ``DFT'' method. To achieve this, for all compounds for which the thermal conductivity was predicted either via the ``Exp.'' or ``Mixed'' method, a comparison is made with what was obtained with the ``DFT'' method. For compounds for which either the ``Exp.'' or ``Mixed'' method was used, and for which experimental or EMD data was available, the raw DFT parameters of the model calculated by Eq. 25 and Eq.27 are reported in Table II along with the adjusted parameters to reproduce simultaneously all or some available experimental data on $C_{P}$, $\alpha_{V}$ and $B_{S}$. These values are also compared with experimental data reported in the literature and determined most of the time regardless of any optimization procedure. DFT calculations were also performed for compounds for which the model parameters were determined via the ``Exp.'' method . For each of these compounds, the Monkhorst Pack grid of dimension is reported in Table II, the plane-Wave kinetic cut-off energy being 520 eV.  The accuracy of the DFT predictions of the lattice thermal conductivity can be quantified through the following relative error function:
\begin{equation}
\Delta_{DFT}^{Exp./Mixed}=1-\dfrac{K^{"Exp."("Mixed")}}{K^{DFT}}
\end{equation}
Neglecting the difference between the temperature dependence of $\delta$ between the  ``Exp.''/``Mixed'' and ``DFT'' methods, the above error function can be approximated as a constant:
\begin{equation}
\Delta_{DFT}^{Exp./Mixed}\simeq 1-\left(\dfrac{\Theta_{D}^{"Exp."("Mixed")}}{\Theta_{D}^{"DFT."}}\right)^{3}  \times \left( \dfrac{\gamma^{"DFT."} }{\gamma^{"Exp."("Mixed")}} \right)^{2}
\end{equation}

 \begin{table}
\centering
\caption{DFT, optimized (Opt.) and experimental (Exp.) values of the two key model parameters, $\Theta_{D}$ and $\gamma$, for the compounds for which either the ``Exp.'' or ``Mixed'' method was employed for the model parametrization. For each compound, the dimension of the Monkhorst Pack grid used to perform the present DFT calculations is also specified, the plane-Wave kinetic cut-off energy being the same for all compounds: 520 eV. $\Delta_{DFT}^{Exp./Mixed}$ defines the ratio (given in $
\%$) between the predict lattice thermal with either method ``Exp.'', ``Mixed'' or ``DFT''.  }

\resizebox{!}{!} {
\begin{tabular}{ c c c c c c c c c c c c c c c c c c}
\hline
Comp.& & &$\Theta_{D}$(K)& & & & $\gamma$ & & $\Delta_{DFT}^{Exp./Mixed}$\\ 
& k-points grid & DFT &  Opt. &  Exp.  & &  DFT &  Opt. & Exp. 

 \\ \hline
NaF&11$\times$11$\times$11& 494 &497&492$\pm$3~\cite{Lewi} &--  & 1.70 & 1.57&1.55~\cite{Bens}&  19.4$\%$	\\
LiF&11$\times$11$\times$11 &690&630&630~\cite{Kreu} &-- &  2.03 & 1.70&1.72~\cite{Rupp}&  8.5$\%$	\\
CaF$_{2}$&11$\times$11$\times$11 &548&495&495$\pm$5~\cite{Huff} &-- &  2.26& 2.10&2.00~\cite{Coll2}&  14.6$\%$		\\
MgF$_{2}$&12$\times$12$\times$16 &627&595&610$\pm$5~\cite{Bark} &-- &  1.50& 1.57&1.61~\cite{Blan2}&  22.0$\%$		\\  
$\beta$--Na$_{3}$AlF$_6$&8$\times$8$\times$8 &508&515&NA &-- &  1.48& 1.51&NA&  0.1$\%$				\\
Na$_{5}$Al$_{3}$F$_{14}$&8$\times$8$\times$4 &511&495&NA &-- &  1.53& 1.58&NA&  14.8$\%$				\\
LiCaAlF$_6$&8$\times$8$\times$4 &525&525&NA &-- &  1.62& 1.48&NA&19.8$\%$		\\
$\alpha$--Al$_{2}$O$_{3}$& 18$\times$18$\times$6 &946&920&1020~\cite{Taru} &-- &  1.67& 1.54&1.66~\cite{Taru}&8.2$\%$	      \\ \hline

\end{tabular} 
}
\label{tab:results}
\end{table} 

\begin{figure}
\centering
\includegraphics[scale=0.5]{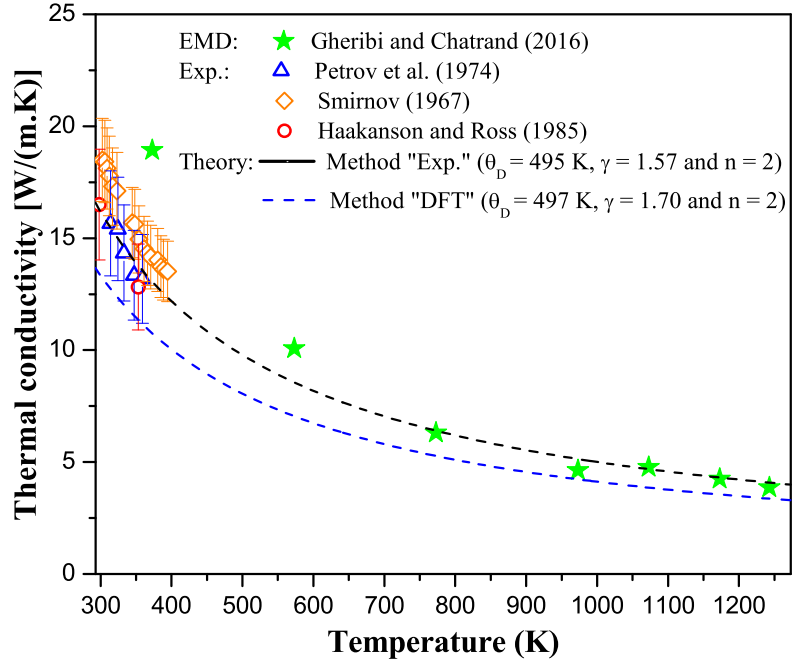} 
\caption{Predicted lattice thermal conductivity of NaF as a function of temperature with the ``Exp.'' method (solid line) and ``DFT'' (dashed line) in comparison with the available experimental data and EMD simulations (solid stars) performed in our previous study~\cite{arxiv}. The experimental data is referenced as follows: Petrov et al. (1974):~\cite{Petr} (open triangles), Smirnov (1967):~\cite{Smi} (open diamonds) and Haakanson and Ross (1985)~\cite{Haak} (open circles).  }
\label{fig2}
\end{figure}
\begin{figure}
\centering
\includegraphics[scale=0.5]{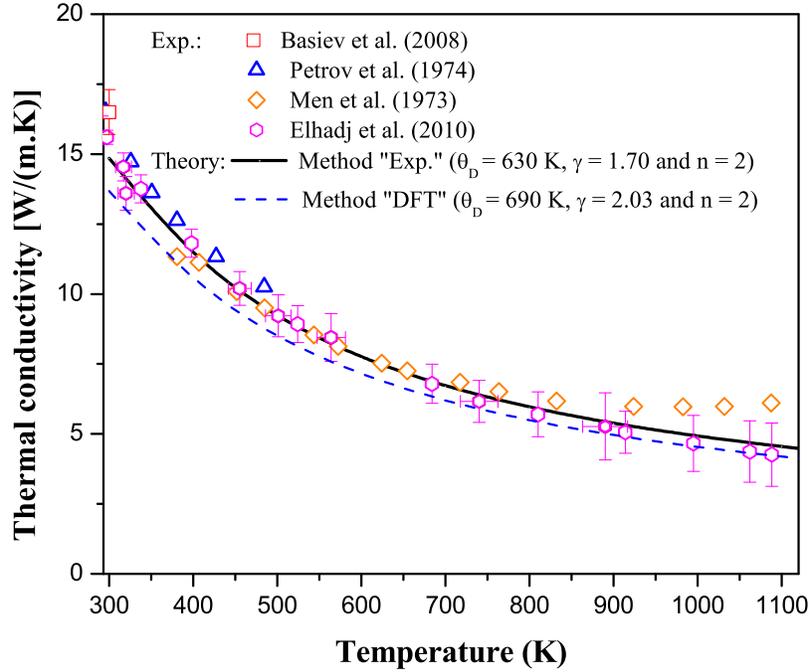} 
\caption{Predicted lattice thermal conductivity of LiF as a function of temperature with the ``Exp.'' method (solid line) and ``DFT'' (dashed line)in comparison with the available experimental which are referenced as follows: Basiev et al. (2008):~\cite{Basi} (open squares), Petrov et al. (1974):~\cite{Petr} (open triangles), Men et al. (1973)~\cite{Men1} (open diamonds) and Elhadj et al. (2010)~\cite{Elha} (open hexagons).  }
\label{fig1a}
\end{figure}
\begin{figure}
\centering
\includegraphics[scale=0.5]{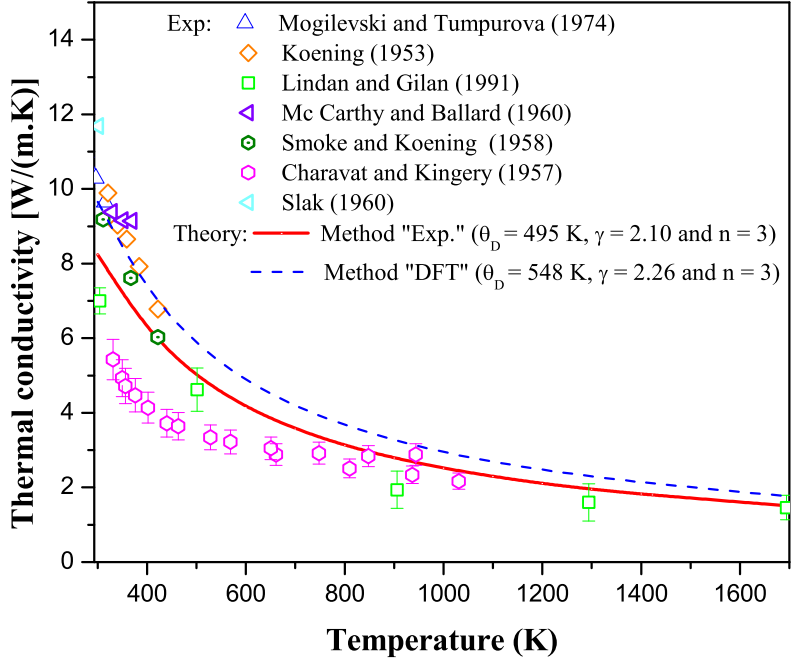} 
\caption{Predicted lattice thermal conductivity of CaF$_{2}$ as a function of temperature with the ``Exp.'' method (solid line) and ``DFT'' (dashed line) in comparison with the available experimental which are referenced in the caption of Figure 1.  }
\label{fig1a}
\end{figure}
\begin{figure}
\centering
\includegraphics[scale=0.5]{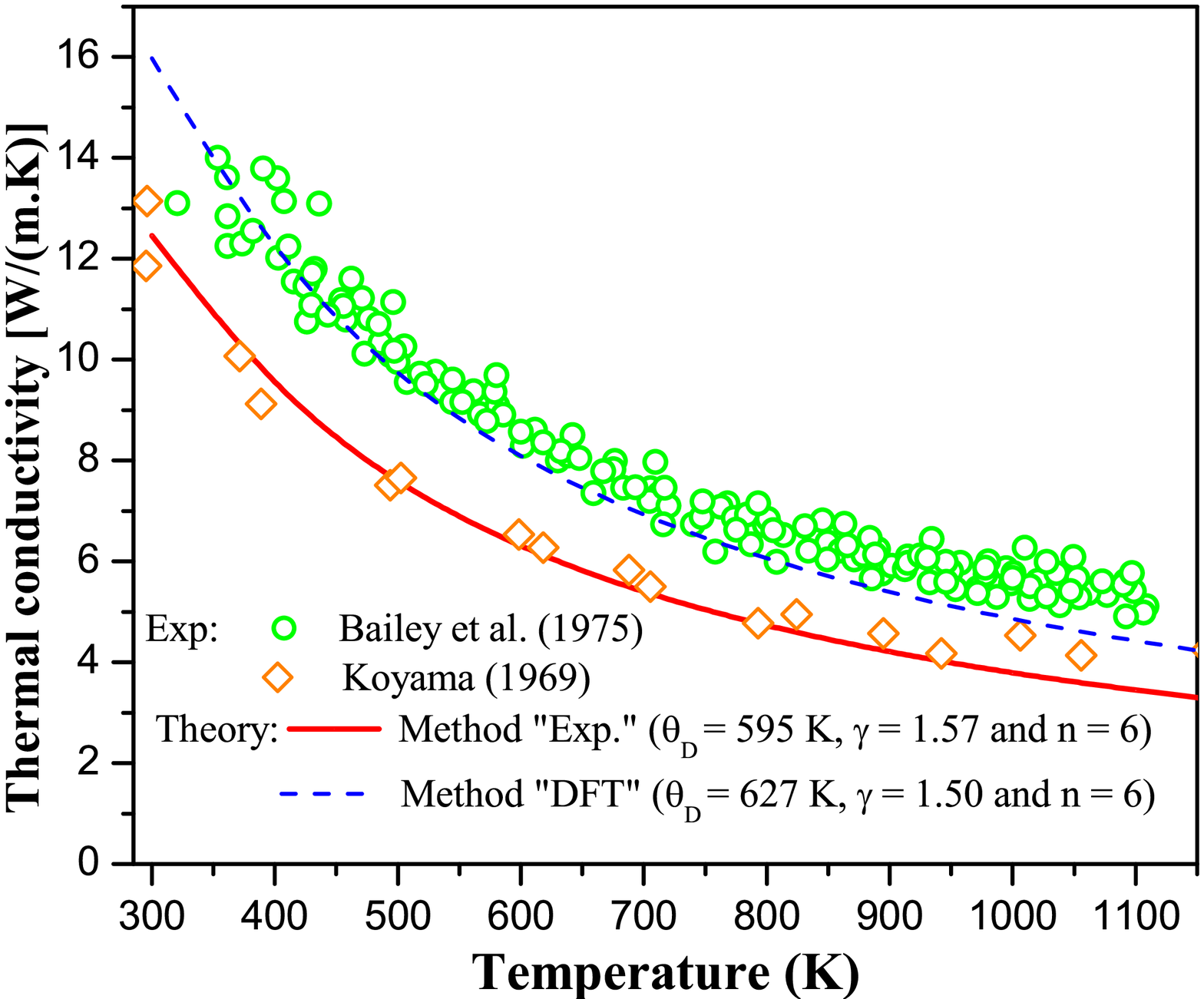} 
\caption{Predicted lattice thermal conductivity of MgF$_{2}$ as a function of temperature with the ``Exp.'' method (solid line) and ``DFT'' (dashed line) in comparison with the available experimental of Bailey et al. (1975):~\cite{Bail2} (open diamonds) and Koyama et al. (1969):~\cite{Koya} (open circles).  }
\label{fig1a}
\end{figure}

With an average error of about $15\%$ and an error of less than $20\%$, the DFT predictions of the lattice thermal conductivity of compounds considered in Table II are in good agreement with those predicted by either the ``DFT'' or ``Exp.'' method. For simple fluoride compounds, NaF, LiF, CaF$_{2}$ and MgF$_{2}$, the predictions of the lattice thermal conductivity with both methods are shown from Figure 2 to Figure 5 in comparison with the available experimental data and, for NaF, with EMD simulation data as well. \\

For NaF (Figure 2), experimental data is available only at low temperature: below 400~K. To alleviate this limitation, EMD simulations were carried out by Gheribi et al.~\cite{arxiv} in order to predict the thermal conductivity of NaF from 400~K up to the melting temperature (1266~K). A good agreement is observed between the predicted thermal conductivity (using the ``Exp.'' method) with experimental data (available only at low temperature) on the one hand and with the EMD simulations above about the Debye temperature one the other hand. This good agreement demonstrates the reliability of both the theoretical predictions using the ``Exp.'' method and EMD simulations for ionic systems above the Debye temperature. The fact that EMD simulations are accurate only above the Debye temperature is extensively discussed in our recent work~\cite{arxiv}. For NaF, the prediction of the thermal conductivity via the ``DFT'' method can be considered to be reasonably accurate as it is within the margin of experimental error and in good agreement with EMD simulations data at high temperature. For LiF (Figure~3), the thermal conductivities predicted by both ``Exp'' and ``DFT'' are in good agreement with the available experimental data. The cases of MgF$_{2}$ and CaF$_{2}$ are more difficult to analyse in terms of accuracy. Indeed, as it can be seen in Figure~4 and Figure~5, the experimental data reported for these two compounds is scattered, the predicted thermal conductivity according to which of the two methods, ``Exp'' or ``DFT'', is employed is in agreement with one (Mgf$_{2}$) or several (CaF$_{2}$) sets of experimental data. While we discuss the accuracy of the methods used to predict the thermal conductivity, it would also be relevant to discuss the reliability of the experimental techniques used to measure the thermal conductivity and thermal diffusivity. Unfortunately this is out of scope of the current work since both methods  predict the thermal conductivity within the margin of the experimental error. 
\begin{figure}
\centering
\includegraphics[scale=0.5]{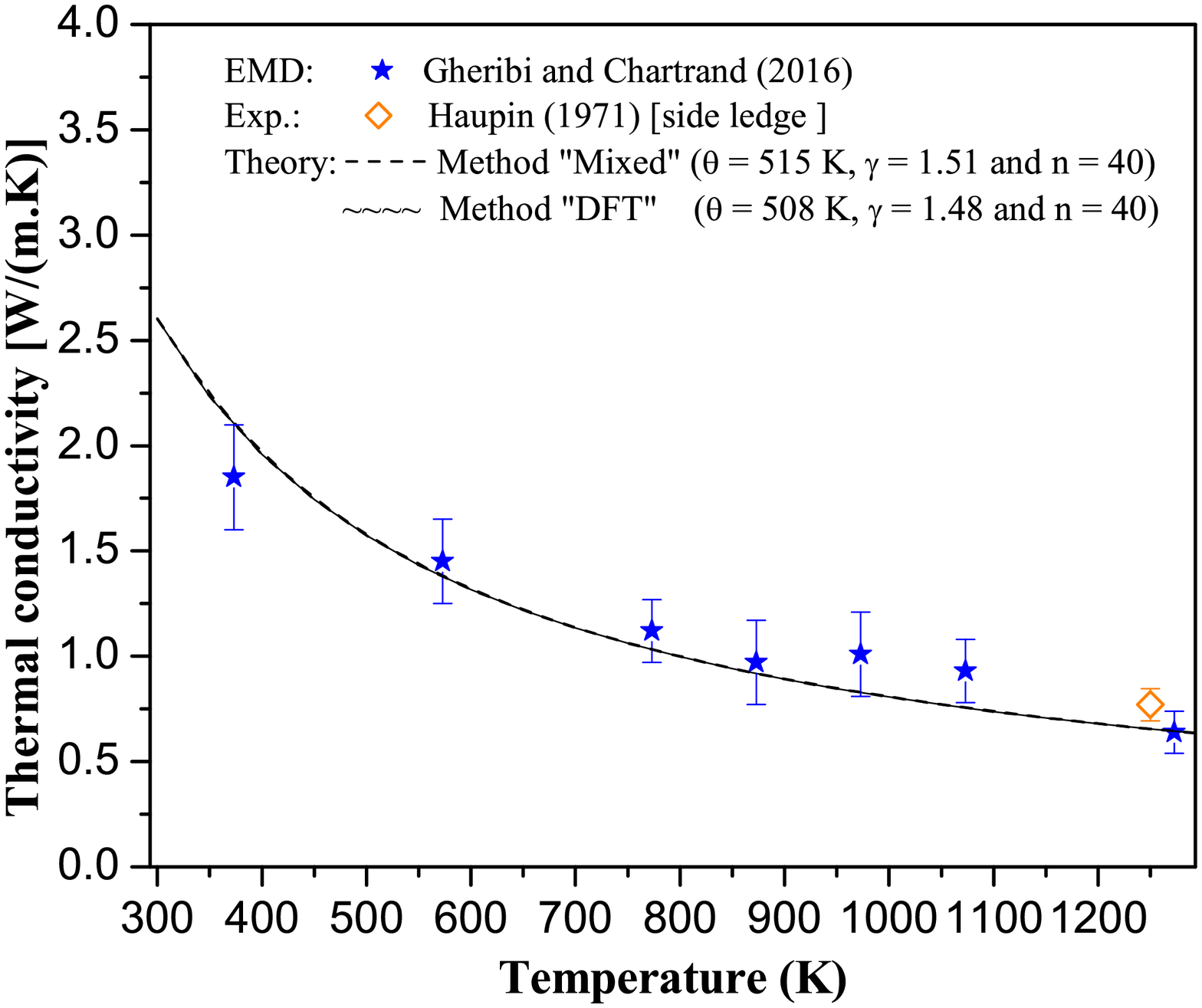} 
\caption{Predicted lattice thermal conductivity of $\beta$--Na$_{3}$AlF$_6$ as a function of temperature using the ``Mixed'' method (solid line) and ``DFT'' (dashed line) in comparison with EMD simulations (solid stars) performed in our previous study~\cite{arxiv} and the experimental data point on side ledge measured by Haupin~\cite{Haup} (open diamond).  }
\label{fig1a}
\end{figure}
\begin{figure}
\centering
\includegraphics[scale=0.5]{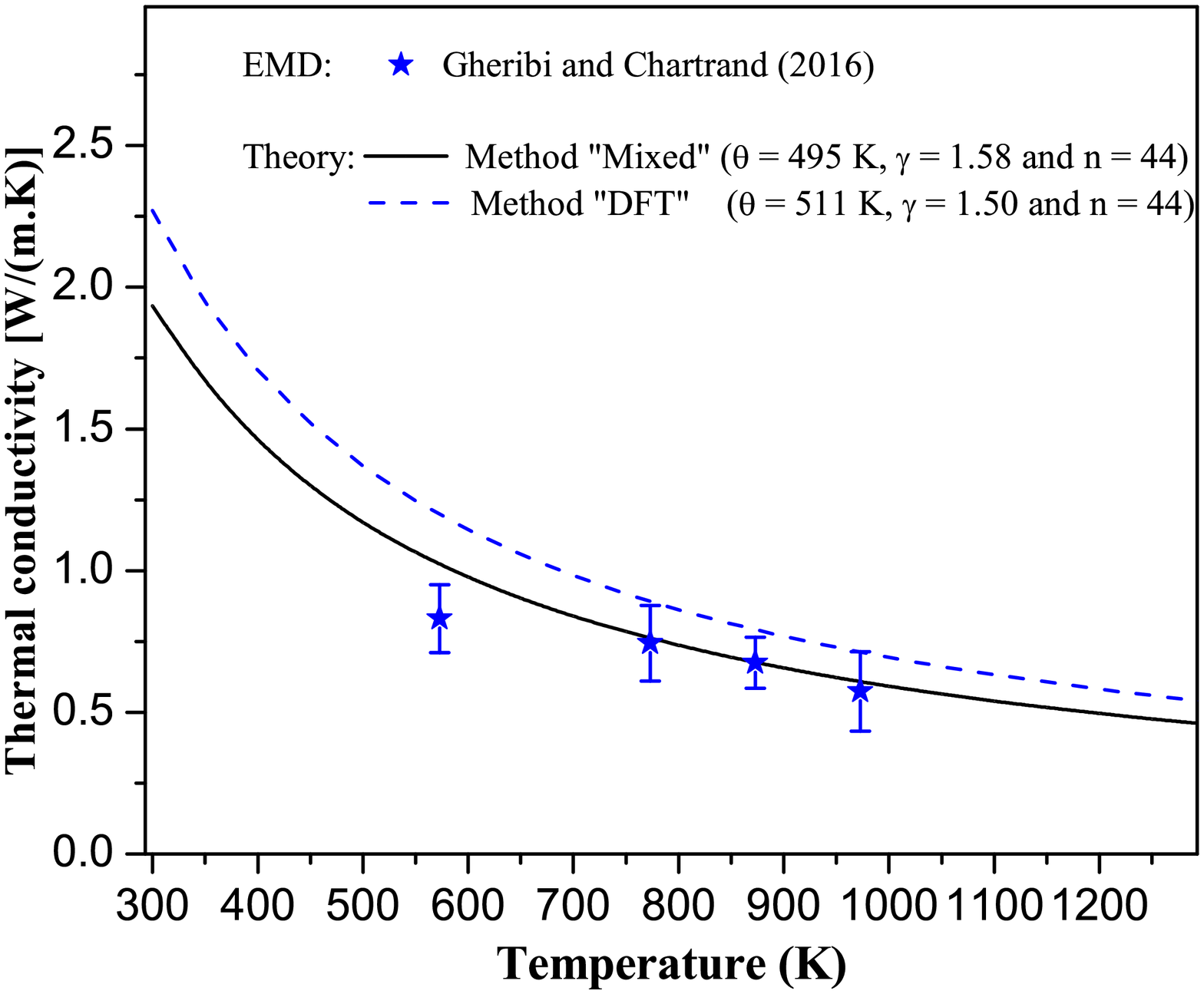} 
\caption{Predicted lattice thermal conductivity of Na$_{5}$Al$_{3}$F$_{14}$ as a function of temperature with method "Mixed" (solid line) and "DFT" (dashed line) in comparison with EMD simulations (solid stars) performed in our previous study~\cite{arxiv}. }
\label{fig1a}
\end{figure}\\

The parametrization of the theoretical model with the ``Mixed'' method has already been presented in our recent work~\cite{arxiv} and will not be repeated here. Briefly, as for the ``Mixed'' method, the theoretical model parameters were first determined by DFT. Then the model parameters were slightly adjusted in order to reproduce simultaneously the experimental data reported on both molar volume and heat capacity as a function of temperature, as shown in Table II. More details on the optimization procedure can be found in~\cite{arxiv}. In Figures~6 and 7 respectively show the predicted thermal conductivity of $\beta$--Na$_{3}$AlF$_6$ and Na$_{5}$Al$_{3}$F$_{14}$ with the ``Mixed'' and ``DFT'' methods. For $\beta$--Na$_{3}$AlF$_6$ both methods are almost identical, this is fortuitous. For these two compounds, no experimental data on thermal transport properties is available. The only experimental data which can be found in the literature is the thermal conductivity of the side ledge close to the liquidus temperature reported by Haupin~\cite{Haup} and given by Eq.~2. The predicted thermal conductivity is in good agreement with both EMD simulations and the Haupin data. Note that contrary to NaF, EMD simulations underestimate the thermal conductivity a low temperature. This point is extensively discussed in~\cite{arxiv}. For Na$_{5}$Al$_{3}$F$_{14}$, the predicted thermal conductivity via the ``Mixed'' method shows also a good agreement with EMD simulations at low temperature due to the same physical reasons than for $\beta$--Na$_{3}$AlF$_6$.\\

\begin{figure}
\centering
\includegraphics[scale=0.5]{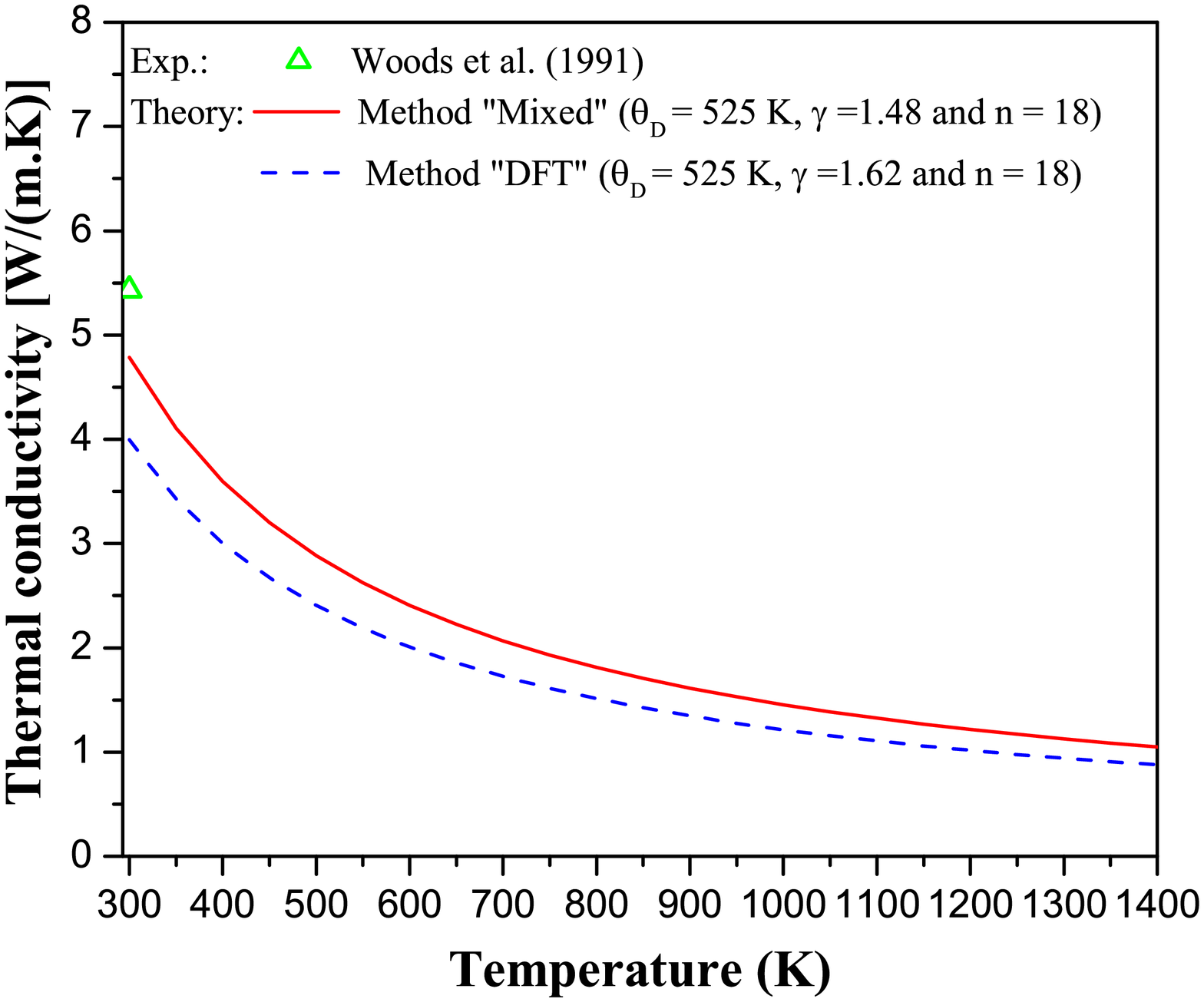} 
\caption{Predicted lattice thermal conductivity of LiCaAlF$_{6}$ as a function of temperature using the ``Mixed'' method (solid line) and ``DFT'' (dashed line) in comparison with the only available experimental data, reported by Woods et al.~\cite{Wood} (open triangle). }
\label{fig1a}
\end{figure}

The thermophysical and thermal transport properties of LiCaAlF$_{6}$ were reported by Woods et al.~\cite{Wood} as a function of temperature for $C_{P}$ and $\alpha_{V}$ and at room temperature for the adiabatic elastic constants. The two components (parallel and perpendicular) of the thermal conductivity and thermal diffusivity tensors are also reported only at room temperature. Basically, the optimization method of the model parameters is similar to that performed for $\beta$--Na$_{3}$AlF$_6$. The parameters were initially determined via ``DFT'' and then adjusted to reproduce the data reported by Woods et al.~\cite{Wood}. The adiabatic bulk modulus predicted was found to be very close to what is reported by Woods et al. For this reason, no adjustment of the Debye temperature was necessary. The Gr\"{u}neisen parameter was adjusted to reproduce simultaneously the thermal expansion and the heat capacity. The predicted thermal conductivities with both the ``Mixed'' and ``DFT'' methods are shown in Figure 8. At room temperature (300~K), the differences between the values reported by Woods et. al.~\cite{Wood} and those predicted by the theoretical model are $18\%$ and $33\%$ respectively for ``Mixed'' and ``DFT''. As no experimental error is provided by Woods et al. and given that there is only one experimental data point reported for LiCaAlF$_{6}$, it can be assumed in this case that the accuracy of the ``Mixed'' method is satisfactory. The rather large error obtained with the ``DFT'' method is mainly due to the over estimation of the Gr\"{u}neisen parameter. \\ 

\begin{figure}
\centering
\includegraphics[scale=0.5]{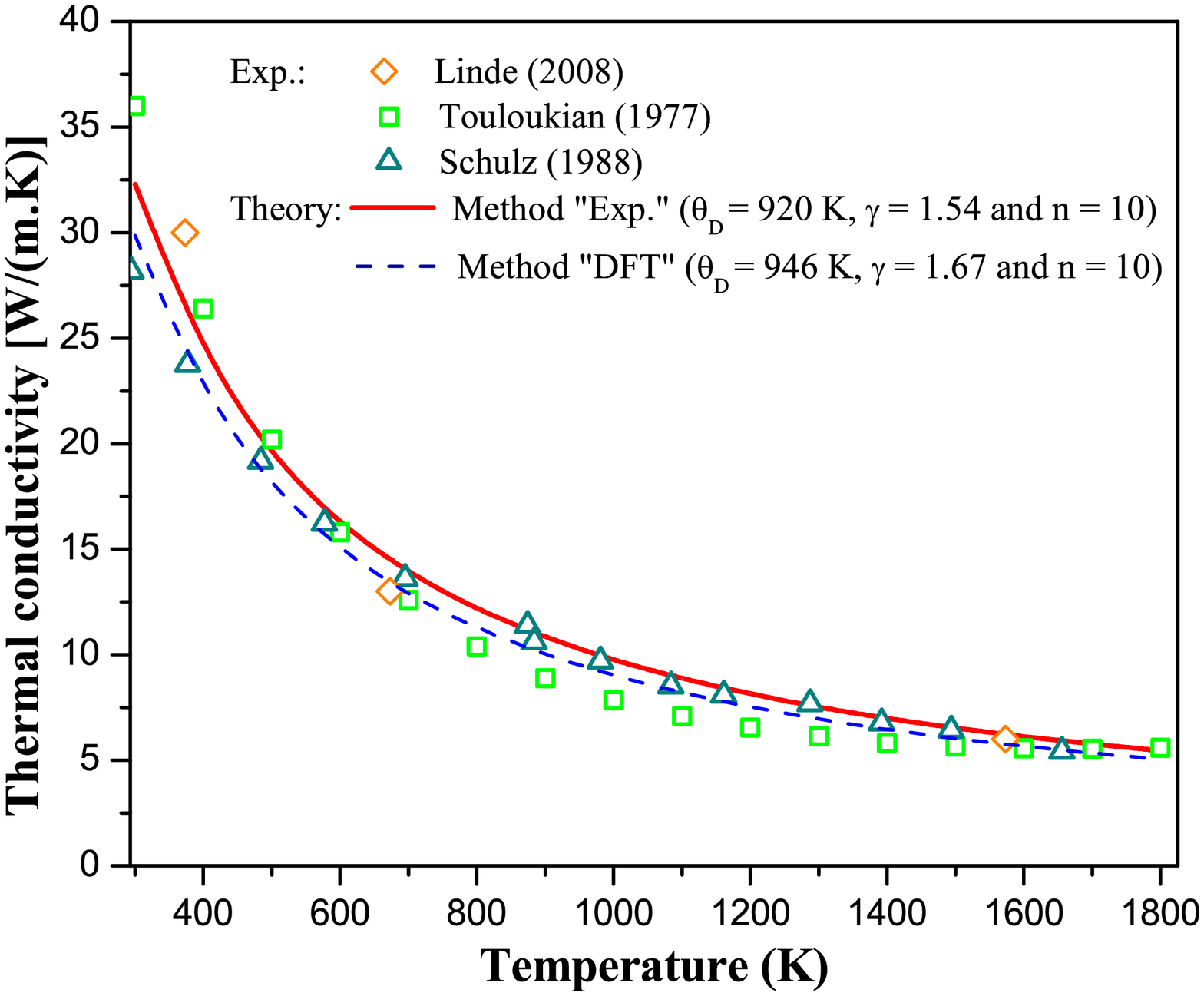} 
\caption{Predicted lattice thermal conductivity of $ \alpha $--Al$_{2}$O$_{3}$ as a function of temperature using the ``Exp.'' method (solid line) and ``DFT'' (dashed line) in comparison with the available experimental data which are referenced as follows: Linde (2008):~\cite{Lide} (open diamonds), Touloukian (1977):~\cite{Toul} (open squares) and Schulz (1988)~\cite{Schu} (open triangles). }
\label{fig1a}
\end{figure}
\begin{figure}
\centering
\includegraphics[scale=0.5]{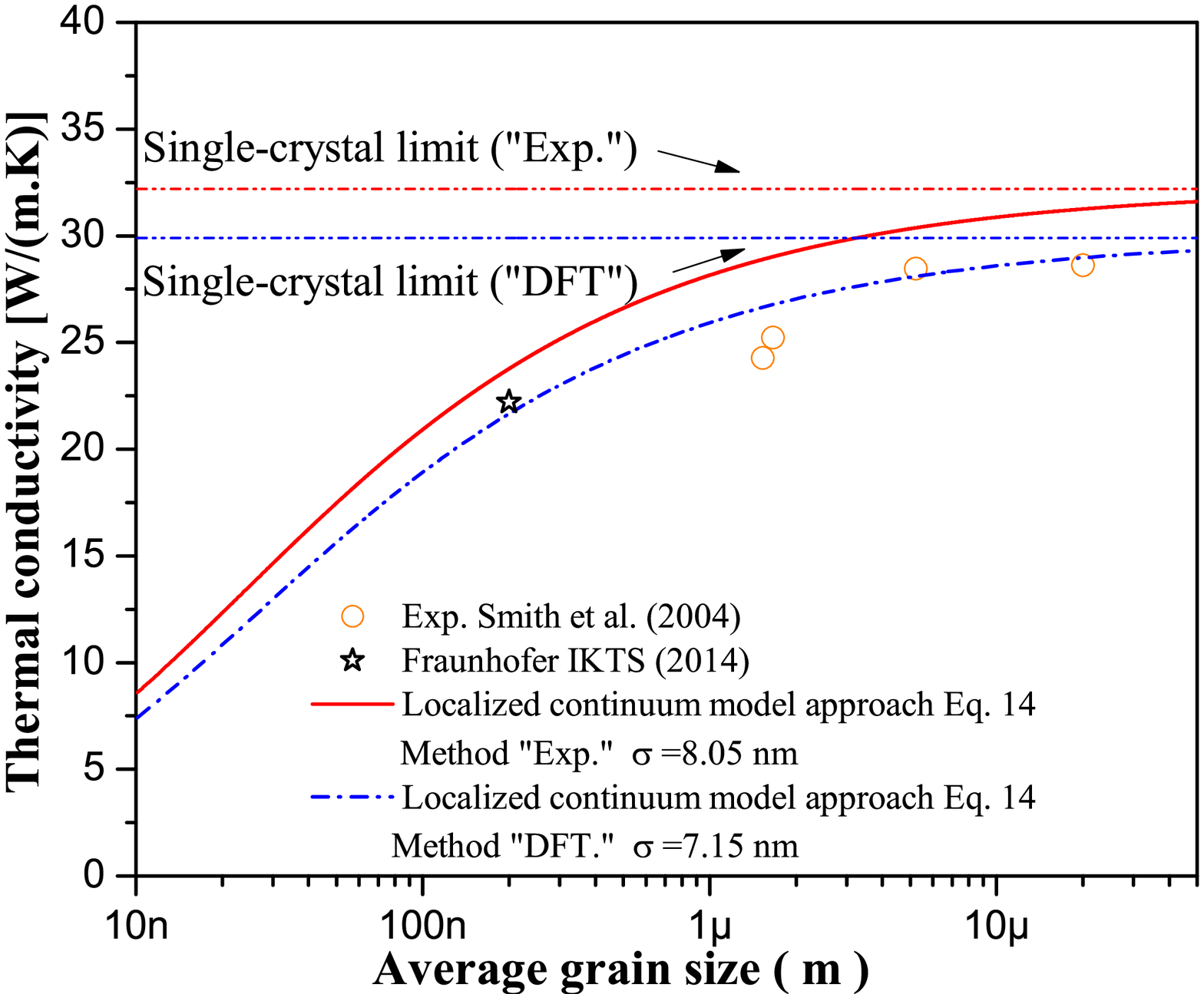} 
\caption{Predicted lattice thermal conductivity of $ \alpha $--Al$_{2}$O$_{3}$ as a function of average grain size using the ``Exp.'' method (solid line) and ``DFT'' (dashed line) in comparison with the available experimental data reported by Smith et al. (2004)\cite{Smi} (open circles) and that of IKTS (2014)~\cite{ikts} (open star). The value of the single-crystal limit obtained respectively via both methods are indicated and represented by dashed dot lines.}
\label{fig1a}
\end{figure}

In the case of $\alpha $--Al$_{2}$O$_{3}$, the optimization of the model parameters by the ``Exp.'' method was already presented in~\cite{Gheribi1}. Details of the optimization procedure will not repeated here. The predicted thermal conductivities via either method, ``Exp.'' or ``DFT'' are in good agreement with the available experimental data, as it is shown in Figure~9. For $\alpha $--Al$_{2}$O$_{3}$ microstructures, some data of thermal conductivity as a function of average grain size  is available. In Figure 10, the predicted thermal conductivity of $\alpha $--Al$_{2}$O$_{3}$ as a function of grain size, obtained by combining Eq. 14 and Eq. 15, is presented for two different characteristic lengths $\sigma$: one calculated with the physical parameters obtained by the ``Exp.'' method and a second one with the raw parameter obtained by ``DFT''. First, note that for a microstructure with an average grain size below 5 $\mu$m, the effect of the grain boundaries can be considered as negligible. At fist glance, the ``DFT'' parametrization seems to be more accurate in predicting the grain size dependence of the thermal conductivity of $\alpha $--Al$_{2}$O$_{3}$. However, the predicted variation of the lattice thermal conductivity with avearage grain size is very close for both methods. The apparent better agreement obtained with the ``DFT'' method lies in the fact that the predicted lattice thermal conductivity of the single crystal (Figure 9) is in better agreement with the experimental data reported by Smith et al.~\cite{Smit} than what is predicted by the ``Exp.'' method, 32.7 W/(m.K) vs. 29.9 W/(m.K). Note that the theroritical model developed by Gheribi and Chartrand~\cite{Gheribi5} fails to predicted the grain size dependence of the lattice thermal conductivity for microstructures with an average grain size below 50 nm. Thus, it is not recommended to use this model for microstructures with nanoscaled grains.
\begin{figure}
\centering
\includegraphics[scale=0.5]{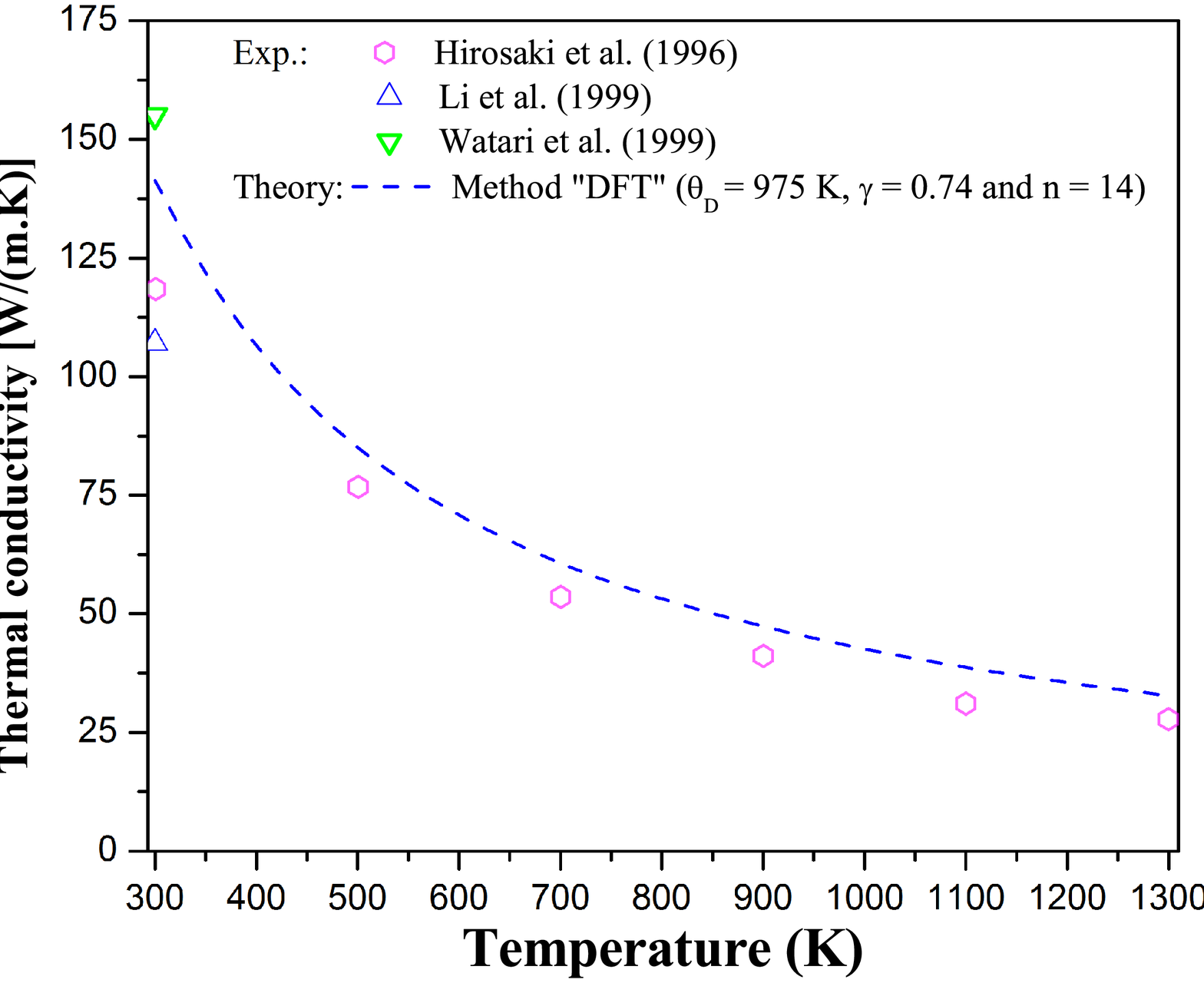} 
\caption{Predicted lattice thermal conductivity of $\beta$--Si${3}$N$_{4}$ as a function of temperature using the ``DFT'' method in comparison with the available experimental data which are referenced as follows: Hirosaki et al. (1996):~\cite{Hiro} (open hexagons), Li et al. (1999)~\cite{Li} (open up triangle) and Watari et al. (1999)~\cite{Wata} (open down-triangle). }
\label{fig1a}
\end{figure}
\begin{figure}
\centering
\includegraphics[scale=0.5]{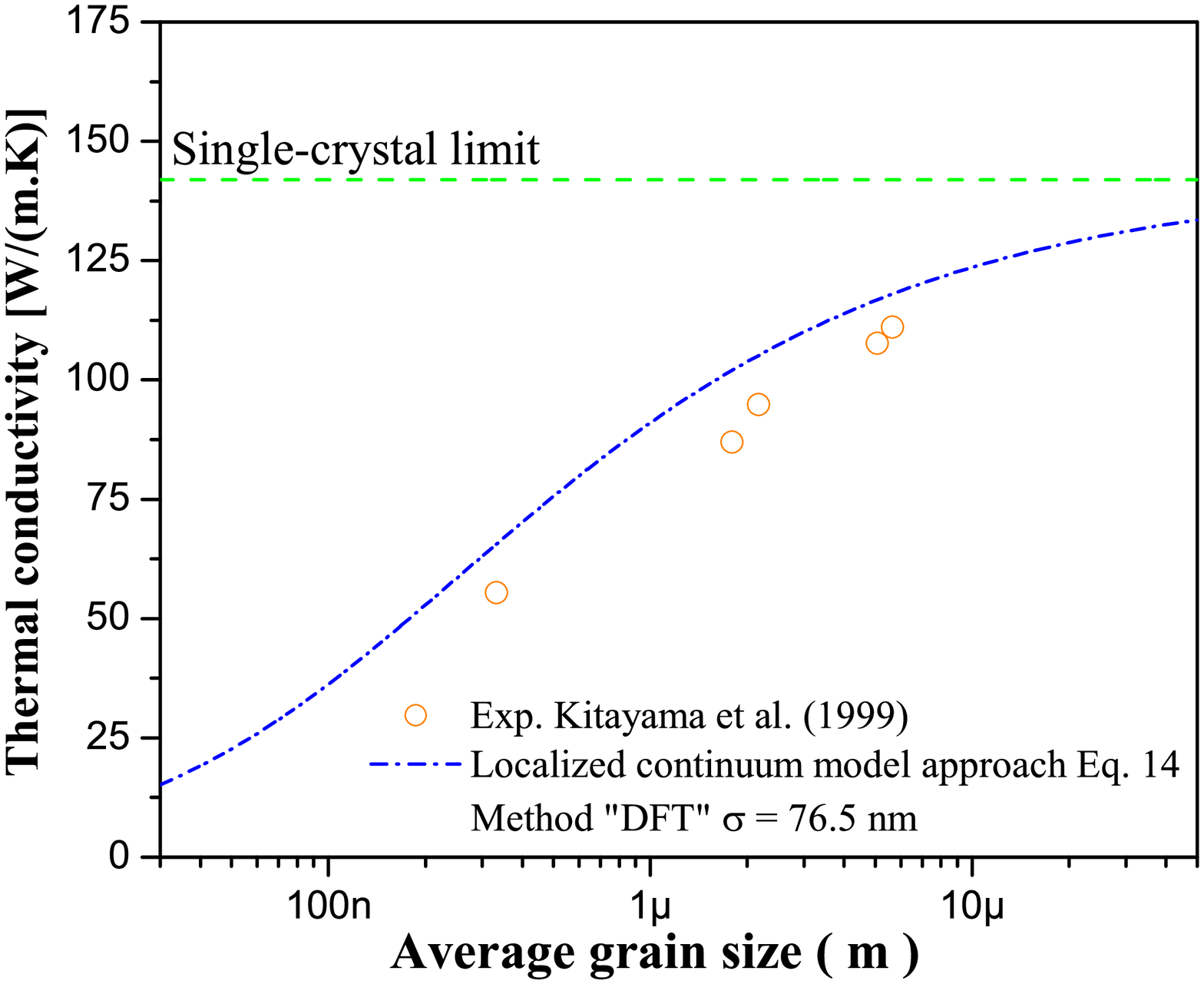} 
\caption{Predicted lattice thermal conductivity of $\beta$--Si${3}$N$_{4}$ as a function of average grain size using the ``DFT'' method in comparison with the experimental set reported by Kitayama et al.~\cite{Kita} (open circles). }
\label{fig1a}
\end{figure}

The last case study for the validation of the ``DFT'' method is $\beta$--Si$_{3}$N$_{4}$. For this compounds, we only consider the ``DFT`` method to predict the thermal conductivity of the single crystal as a function of temperature and the thermal conductivity of microstructure as a function of average grain size at room temperature~(300K). The results are shown in Figure 11 and Figure 12 in comparison with the available experimental data. At room temperature, the experimental data for the single crystal lies in the range of 106-155~W/(m.K) while our predictions give~142 W/(m.K). At high temperature, the predicted thermal conductivity is in good agreement with the experimental data reported by Hirosaki et al.~\cite{Hiro} with less than 25$\%$ difference. Given the large dispersion of the experimental data observed at 300~K, it can be reasonably assumed that the predicted thermal conductivity of $\beta$--Si$_{3}$N$_{4}$ is accurate.\\

The grain size dependence of the lattice thermal conductivity is also predicted with a good accuracy (Figure 12) as it matches with the experimental data reported by Kitayama et al.~\cite{Kita} with an error less than 20 $\%$.\\

For Al$_{4}$C$_{3}$, no experimental data on thermal transport properties is available in the literature. The ``Mixed'' method was employed to determined the model parameters. Only few experimental data is available, namely the heat capacity measured from 0 to 2000~K by Furukawa et al.~\cite{Furu} and the equation of state (providing $B_{S}$) measured at 30~K and up to 6~GPa by Solohenko and Kurakevych~\cite{Soloz}. These two sets of data were considered to adjust the ``DFT'' values of $\rho_{298}$, $\Theta_{D}$ and $\gamma$. 

\section{Conclusion and perspective}

The control of the side ledge thickness is essential to maintain a reasonable lifetime of the aluminium electrolysis cells. The monitoring of the side ledge thickness by numerical modelling requires an accurate expression of the thermal conductivity as a function of temperature. The major limitation of the numerical modelling of the side ledge thickness is the lack of experimental data for the side ledge microstructures and phases present in the side ledge.\\

To alleviate this severe lack of data, we have carried out a theoretical study to predict the thermal conductivity of all the compounds which are potentially present in the side ledge. To achieve this, a reliable physical model was considered. This model links the density of the lattice vibration energy and the phonon mean free path to key parameters: the high temperature limit of the Debye temperature and the G\"{u}neisen constant.
Three different types of parametrization for the model were considered: (i) a parametrization based on available experimental data on heat capacity, thermal expansion and adiabatic bulk modulus, (ii) a purely predictive parametrization based on Density Functional Theory (\textit{ab initio}), and (iii) a mixed methodology when few experimental data is available. In the latter case, the parameters are at first determined via DFT calculations and then they are adjusted to reproduce simultaneously all the available experimental data. For compounds for which experimental or equilibrium molecular dynamics (EMD) were available, it was found that the three methodologies give rather accurate predictions of thermal conductivity as a function of temperature, although the two methods based on experimental data provide more accurate predictions.  The extension of the model for the description of the grain size dependence of the thermal conductivity  microstructures was found also accurate for two compounds for which data is available, namely $\alpha$--Al$_{3}$O$_{3}$ and $\beta$--Si$_{3}$N$_{4}$. Finally, a database for thermal conductivity and density as a function of temperature is provided for all compounds potentially present in the side ledge.\\
    
Experimental work was recently carried out in order to determine the thermal conductivity of several side ledge samples coming from industrial aluminium electrolyses cells (Rio Tinto Alcan). In a subsequent paper, the experimental results will be presented along with a model for the thermal conductivity of microstructures. This model is based on Eq.~3 with (i) a suitable function $\psi$ , (ii) a reliable model for the porosity dependence~\cite{Garda, Garda2, Garda3} and (iii) the present model (Eq.~14 and Eq.~15) for describing the average grain size dependence. The aim being to provide an accurate model for the prediction of the thermal transport properties of the side ledge as a function of temperature, phases distribution and the minimum microstructure parameters: type of microstructure, grain size and porosity. 

\subsection*{Acknowledgement}
This research was supported by funds from the Natural Sciences and Engineering Research Council of Canada (NSERC) and Rio Tinto Alcan. Computations were made on the supercomputer Briar\'{e} at the Universit\'{e} de Montr\'{e}al, managed by Calcul-Qu\'{e}bec and Compute Canada. The operations of this supercomputer is funded by the Canada Foundation for Innovation (CFI), NanoQu\'{e}bec, RMGA and the Fonds de recherche du Québec - Nature et technologies (FRQ-NT). Eve B\'{e}lisle is warmly acknowledged for all her help and assistance during the writing of this paper. 

%\clearpage

\bibliography{References}

%\printbibliography

\end{document}